\documentclass[iop]{emulateapj}
\usepackage[colorlinks,urlcolor=blue,citecolor=black,linkcolor=blue]{hyperref}
\usepackage{amsmath}

\newcommand{\h}{\ensuremath{{\rm H}}}
\newcommand{\water}{\ensuremath{{\rm H_2O}}}
\newcommand{\hm}{\ensuremath{{\rm H}_2}}
\newcommand{\OH}{\ensuremath{{\rm OH}}}
\newcommand{\xel}{\ensuremath{x_{\rm e}}}
\newcommand{\nel}{\ensuremath{n_{\rm e}}}
\newcommand{\mh}{\ensuremath{m_{\rm H}}}
\newcommand{\mH}{\ensuremath{m_{\rm H}}}
\newcommand{\xh}{\ensuremath{x({\rm H})}}
\newcommand{\nh}{\ensuremath{n_{\rm H}}}
\newcommand{\nH}{\ensuremath{n_{\rm H}}}
\newcommand{\Nh}{\ensuremath{N_{\rm H}}}
\newcommand{\xHe}{\ensuremath{x_{\rm He}}}
\newcommand{\xHH}{\ensuremath{x_{{\rm H}_2}}}
\newcommand{\xhm}{\ensuremath{x({\rm H}_2)}}
\newcommand{\Td}{\ensuremath{T_{\rm d}}}
\newcommand{\Tg}{\ensuremath{T_{\rm g}}}
\newcommand{\ag}{\ensuremath{a_{\rm g}}}
\newcommand{\nd}{\ensuremath{n_{\rm d}}}
\newcommand{\Sd}{\ensuremath{S_{\rm d}}}
\newcommand{\rhod}{\ensuremath{\rho_{\rm d}}}
\newcommand{\rhog}{\ensuremath{\rho_{\rm g}}}

\newcommand{\taud}{\ensuremath{\tau_{\rm d}}}
\newcommand{\Qabs}{\ensuremath{Q_{\rm abs}}}

\newcommand{\Lya}{\ensuremath{{\rm Ly}\alpha}}
\newcommand{\ph}{\ensuremath{_{\rm phel}}}
\newcommand{\phel}{\ensuremath{_{\rm phel}}}
 
\newcommand{\LFUV}{\ensuremath{L_{\mathrm{FUV}}}} 

\newcommand{\FFUV}{\ensuremath{F_\mathrm{FUV}}}
\newcommand{\Lx}{\ensuremath{L_{\rm X}}}
\newcommand{\Tx}{\ensuremath{T_{\rm X}}}

\newcommand{\ah}{\ensuremath{\alpha_{\rm h}}}

\newcommand{\Lsun}{\ensuremath{\,L_{\odot}}}
\newcommand{\LSun}{\ensuremath{\,L_{\odot}}}
\newcommand{\Msun}{\ensuremath{\,M_{\odot}}}
\newcommand{\MSun}{\ensuremath{\,M_{\odot}}}

\newcommand{\RSun}{\ensuremath{\,R_{\odot}}}
\newcommand{\erg}{\ensuremath{{\rm erg}}}
\newcommand{\ergps}{\ensuremath{{\rm erg}\,{\rm s}^{-1}}}
\newcommand{\sqcm}{\ensuremath{{\rm \,cm}^2}}
\newcommand{\psqcm}{\ensuremath{{\rm cm}^{-2}}}
\newcommand{\ps}{\ensuremath{{\rm s}^{-1}}}
\newcommand{\gpcc}{\ensuremath{\rm \,g\,cm^{-3}}}
\newcommand{\pcc}{\ensuremath{{\rm \,cm}^{-3}}}
\newcommand{\pyr}{\ensuremath{{\rm yr}^{-1}}}
\newcommand{\be}{\begin{equation}}
\newcommand{\ee}{\end{equation}}

\shorttitle{WATER AND OH IN PROTOPLANETARY DISKS}
\shortauthors{\'AD\'AMKOVICS, GLASSGOLD \& NAJITA}

\begin{document}
\title{Shielding by Water and OH in FUV and X-ray Irradiated Protoplanetary Disks}

\author{M\'at\'e \'Ad\'amkovics$^1$, Alfred E. Glassgold$^1$, and Joan R. Najita$^2$}
\affil{$^1$Astronomy Department, University of California, Berkeley, CA 94720;
mate@berkeley.edu and aglassgold@berkeley.edu \\
$^2$National Optical Astronomy Observatory, 950 North Cherry Avenue, Tucson, AZ 85719;
 najita@noao.edu
}

\begin{abstract}
We present an integrated thermal-chemical model for the atmosphere
of the inner region of a protoplanetary disk that includes irradiation
by both far ultraviolet (FUV) and X-ray radiation. 
We focus on how the photodissociation of \water\ and OH affects the abundances
of these and related species and how it contributes to the heating
of the atmosphere. The dust in the atmosphere plays several important
roles, primarily as the site of H$_2$ formation and by absorbing
the FUV. Large amounts of water can be synthesized within the inner 4\,AU 
of a disk around a typical classical T~Tauri star. 
OH is found primarily at the top
of a warm region where the gas temperature is $\Tg \approx 650 - 1000$\,K and \water\ below
it where the temperature is lower, $\Tg \approx 250 - 650$\,K. The
amounts of \water\ and OH and the temperatures of the regions in which they formed 
are in agreement with recent {\it Spitzer} measurements and support the notion of
{\it in situ} production of water in the inner regions of protoplanetary
disks.  
We find that the synthesized water is effective in shielding the disk
midplane from stellar FUV radiation.

\vspace{1cm}
\end{abstract}

\keywords{astrochemistry --- planetary systems: protoplanetary disks --- X-rays: stars}

\section{Introduction}

The gas and dust in protoplanetary disks are both important in the formation of stars and planets \citep[e.g.,][]{Dullemond2010, Williams2011}. Significant progress has been made in recent years by observing the gas with ground-based facilities and the {\it Spitzer} and {\it Herschel} space observatories. A variety of molecules have been detected in the inner planet forming regions of these disks, notably water and simple organic molecules \citep{Carr2004,CN08,CN11, Pascucci2009,Pontoppidan2010a,Pontoppidan2010b,Salyk2011,Najita2013}. The presence of water is of particular astrophysical and astrobiological significance. Water can be as abundant as CO in protoplanetary disks, making it the third most abundant molecule after H$_2$ and CO. Water and dust can also play important physical and chemical roles, a focus of this article.

Numerous models of protoplanetary disk chemistry have been developed for the purpose of establishing and analyzing diagnostic spectral lines of the gas \citep{Markwick2002, Kamp2004, Nomura2005, Nomura2007, Nomura2009, Agundez2008, Gorti2008, Woods2009, Woitke2009a, Woitke2009b, Kamp2010, Fogel2011, Kamp2011, Thi2010,Heinzeller2011, Walsh2012, Aresu2011, Aresu2012, Bruderer2012, Meijerink2012, Akimkin2013}. These models tend to differ in significant physical details, e.g., how the gas is irradiated and how it is heated, as well as in the chemistry. In addition to our earlier paper \citep{GMN09}, several other models produce significant levels of water in the inner disk \citep[e.g.,][]{Woods2009, Thi2010, Fogel2011, Heinzeller2011, Aresu2011}, and thus support the idea of {\it in situ} formation in the gas phase. If this conclusion can be strengthened, as we attempt to do here, it would diminish the need for an origin of water from the transport of icy particles and bodies \citep[e.g.,\!\!~][]{Ciesla2006}.   

In 2009 Bethell and Bergin called attention to the possibility that the disk midplane is shielded from dissociating far ultraviolet (FUV) radiation by water and OH in the disk atmosphere \citep[henceforth BB09]{BB09}. They described how including such molecular shielding could account for several features of the water and OH emission observed by {\it Spitzer}. Molecular shielding becomes important when grains have settled out of the atmosphere and dust is no longer the dominant absorber of UV radiation. Shielding by molecules can limit the penetration of the FUV through the disk atmosphere and thus affect the transition from regions dominated by atomic oxygen to regions with oxygen-bearing molecules. 

To make these points BB09 adopted an FUV irradiated isothermal atmosphere
and employed a simplified chemistry for OH and \water. 
In particular, they assumed hydrogen to be completely in molecular form. 
One potential difficulty with this assumption is that the formation of \water\ and OH requires H$_2$ as a precursor, and the formation of H$_2$ on grains may be limited when the grain surface area is reduced by grain settling.
A more complete thermal-chemical model is, therefore, needed to explore how well shielding by \water\ and OH works when when grain settling is advanced. This topic has not been pursued by  
earlier thermal-chemical models. For example, the protoplanetary disk model (ProDiMo) of Woitke et al.\ (2009; see also Aresu et al. 2011, Meijerink et al. 2012), while detailed and inclusive of many thermal and chemical processes, includes molecular shielding by H$_2$ and CO, but not \water\ and OH. 

In this paper we  incorporate UV irradiation and molecular shielding by \water\ and OH into our earlier thermal-chemical model of the inner disk atmosphere to explore role of \water\ and OH as UV opacity sources.
We treat carefully a number of physical processes related to the abundance of water and OH.
Unlike BB09, who prescribed two isothermal layers, we evaluate the temperatures in the 
transition between these two regions of the atmosphere by considering the chemistry-dependent thermal rate equations. 
In our  previous study of water with a purely X-ray model \citep{GMN09}, we obtained significant levels of water in the inner regions of a typical T Tauri star disk, but much lower levels of OH than are found by {\it Spitzer} (NAG11) or in the model of BB09. 
Therefore, we also explore in this paper whether a simple UV-irradiated thermal-chemical model can  account for the properties of both the \water\ and OH emission detected with {\it Spitzer}. 

In the next section we describe how FUV photodissociation is treated in our model, including how it adds to the heating of the disk atmosphere. One of our goals is to describe our assumptions and the underlying reasons for them in enough detail to allow useful comparison with the results of other investigations. In Section 2, we describe improvements to our model, most importantly the addition of UV photodissociation and its associated heating. We have also added photoelectric and H$_2$ formation heating to our model. 

\section{Thermal-Chemical Irradiation Model}

The disk model that we irradiate with both X-rays and FUV is derived from earlier versions with only X-rays (\citeauthor{GNI04}~2004, henceforth GNI04; \citeauthor{GMN09}~2009, henceforth GMN09; \citeauthor{NAG11}~2011, henceforth NAG11; \citeauthor{AGM11}~2011, henceforth AGM11). The underlying disk structure is a hydrostatic equilibrium calculation by \citet{DAlessio1999} for a uniform mixture of dust and gas in Keplerian rotation heated by star light and by accretion. GNI04 focused on the gas component separately by including gas cooling and heating (including X-rays) and a chemical model appropriate to the disk atmosphere at small radial distances.
In subsequent papers, GMN09 showed how \water\ could be formed {\em in situ} in inner disk atmospheres if H$_2$ forms on warm grains \citep{Cazaux2004}. NAG11 used this model to study the formation of hydrocarbons in inner disk atmospheres. 
In the present paper, we extend our model by including FUV irradiation and the results of an improved treatment of X-ray ionization and heating (AGM11). In this section, we first present a detailed discussion of the photodissociation of \water\ and OH and then describe several improvements to
the thermal part of the model. The present treatment of photodissociation can be extended to other molecules and we will report on these applications in a future publication.

\subsection{Photodissociation}

\begin{figure}[t]\begin{center}
\includegraphics[width=3.25in]{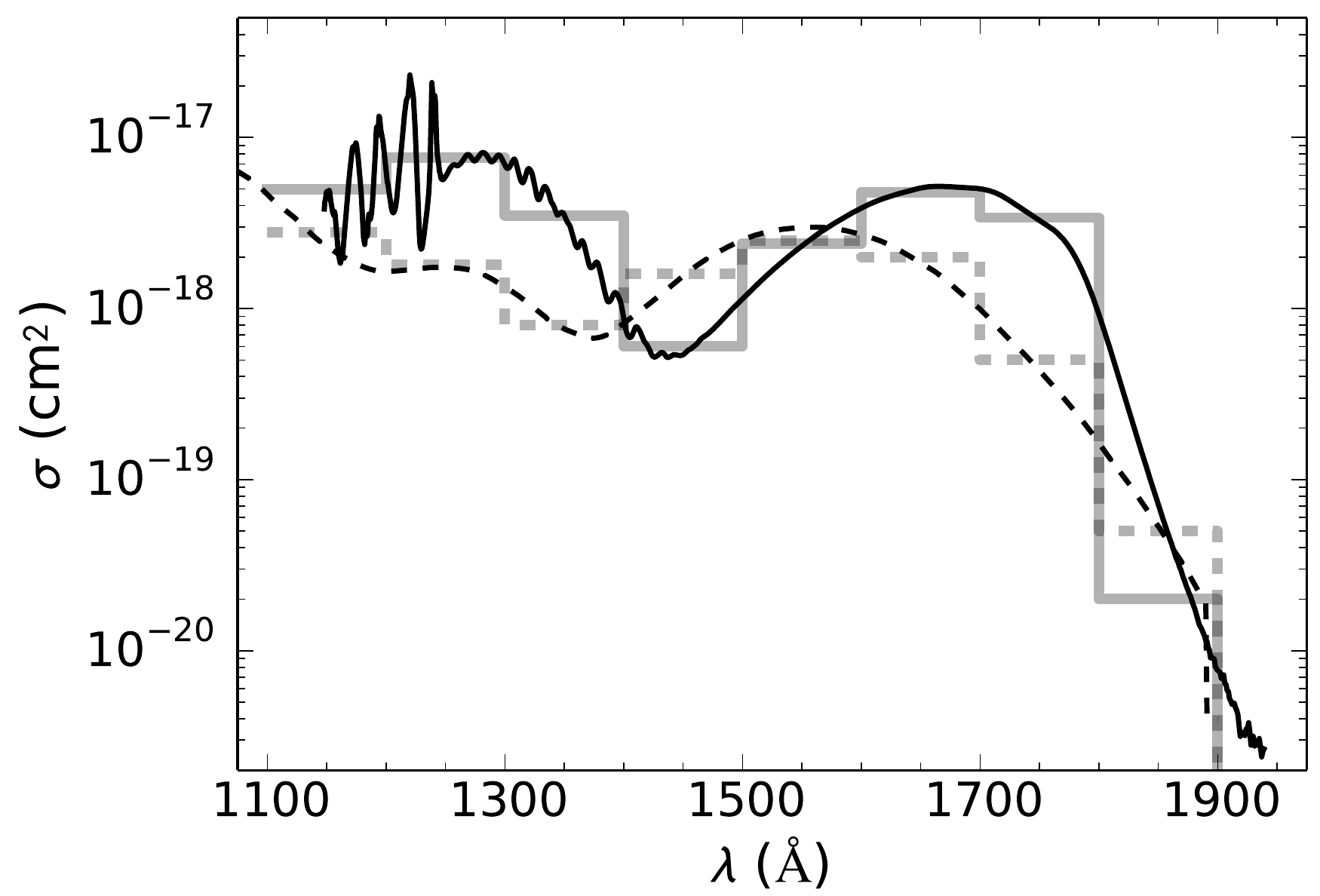}
\caption{\label{fig:sigma_FUV}FUV absorption cross sections, $\sigma(\lambda)$, for water (solid) and OH (dashed) from the literature. The mean values over 100\,\AA\, bins (in thick grey lines) illustrate the values used here for the calculation of photodissociation rates.}
\end{center} \end{figure}

The photodissociation rate for a molecule at a particular radial distance $r$ and altitude $z$ above the disk midplane is given by,
\be
\label{phrate}
G(r,z) = \int_{\lambda_0}^{\lambda_f}  F(\lambda;r,z) \sigma (\lambda) d\lambda \,,
\ee
where $F(\lambda;r,z)$ is the spectral distribution of the local FUV photon number flux in $\psqcm\,\ps$\,\AA$^{-1}$ over the wavelength range from $\lambda_0$ to $\lambda_f$ in\,\AA, and $\sigma(\lambda)$ is the photo-absorption cross section in $\mathrm{cm^{2}}$. There are extensive laboratory measurements of the \water\ cross section, and we use the high-resolution results presented in Figure~1 of \citet{Mota2005}, displayed here as the solid line in Figure~\ref{fig:sigma_FUV}. 
For OH we use the theoretical results from Figure~2 of \citet{vDD84}, displayed as the dashed line in 
Figure~\ref{fig:sigma_FUV}. 
\break
They are in rough agreement with measurements at long wavelengths \citep{NeeLee1984}, but are likely to be less accurate at wavelengths below 1300\,\AA\ due to the coupling of excited states and the occurrence of resonances. The OH photo-absorption cross section is less accurate than that for \water, and it may be substantially revised (probably upwards) when more definitive experiments become available. We average the cross sections over nine 100\,\AA\ bands, and display them as the thick grey lines in Figure~\ref{fig:sigma_FUV}. We ignore the effects on OH and \water\ of FUV radiation below 1100\,\AA\ due to strong line self-shielding by abundant CO and H$_2$ of the radiation from 911.7--1108\,\AA. 

We calculate $F(\lambda;r,z)$ from measurements of the FUV luminosity, $L_{\mathrm{FUV}}$, of T Tauri stars obtained with the {\em Hubble Space Telescope} ({\it HST}) by \citet{Yang2012},
\be
\label{fuv_flux}
F(\lambda;r,z) =  \frac{L_{\mathrm{FUV}}}{4\pi (r^2+z^2)} \,
		    \frac{\lambda}{hc} \, 
		    e^{-( \taud + \tau_{\water} + \tau_{\OH} )},
\ee 
where $\taud$, $\tau_{\water}$ and $\tau_\OH$, are the extinctions due to dust, \water, and OH, respectively. 
The measured luminosities imply a value of $F(\lambda;r,z)$ at $\lambda$=1000\,\AA\, and $r$=1\,AU for a classical T Tauri star (CTTS) that is ten million times larger 
than the interstellar radiation field \citep{Habing1968}.

\citet{Yang2012} report extinction-corrected FUV luminosities for CTTS between 1250 and 1700\,\AA\ excluding \Lya. We adopted 
$\LFUV=0.013\,\Lsun = 5 \times 10^{31}\,$erg~s$^{-1}$, a useful value for comparison with the results of {\em Spitzer} spectroscopy, as molecular emission has been reported for sources with both larger and smaller $\LFUV$ \citep{CN11}. 
This value corresponds to a mean luminosity per 100\,\AA\ of 
$L_{\mathrm{band}}$ = 1.1$\times 10^{31}\,$erg~s$^{-1}$ when taking into account the 450\,\AA\ bandpass of the measurements. In the absence of equivalent measurements outside the 1250 to 1700\,\AA\ range, we use the same $L_{\mathrm{band}}$ in all nine 100\,\AA\ bands from 1100 to 2000\,\AA\ for a total FUV luminosity of 0.025 \LSun. 
This value is similar to but smaller than the smallest FUV luminosity used by BB09 (0.1 \Lsun), who assume that the FUV luminosity includes a large contribution from \Lya\ radiation. 

The \Lya\ line is estimated to make up 70-90\% of the FUV luminosity of CTTS \citep{Schindhelm2012}. 
Since we do not include \Lya\ irradiation in our models, because of its complicated radiative transfer (i.e., H {\sc i} scattering), our adopted reference FUV luminosity underestimates the FUV incident on the disk by a factor of 3-10. 
One way to account roughly for the \Lya\ luminosity of the source would be to assume that our reference FUV luminosity represents 3-10 times the continuum \LFUV\ values. In that case, our adopted reference FUV luminosity of 
0.025\,\LSun\ would still be relevant to CTTS studied with {\em Spitzer} because of the broad \LFUV\ distribution of CTTS (Yang et al.\ 2012).

Consistent with the focus on the photodissociation of \water\ and OH, we add just two reactions to the chemical network used in AGM11,
\be \label{R:phH2O}
\water\ + h\nu  \xrightarrow{}  \rm{OH} +  \rm{H} \tag{R1}
\ee and \be \label{R:phOH}
{\rm OH} + h\nu      \xrightarrow{}  {\rm O  + H}.   \tag{R2} 
\ee
We ignore the role of these reactions in producing excited states of OH and O.

The opacity due to molecular extinction is given by 
\be
\label{molopdepth}
\tau_i = \sigma_i(\lambda)\,N_{\rm los}(i),
\ee
where $N_{\rm los}(i)$ is the column density from the star for molecule $i$ ($i$ = \water\ or OH). The calculations of $N_{\rm los}(i)$ are non-trivial because they depend on the molecular abundances throughout the disk, which themselves depend on column density through Eq.~\ref{fuv_flux} and the photochemical reactions \ref{R:phH2O} and \ref{R:phOH}. In our previous X-ray models, this complication was unimportant because the X-ray intensity is nearly independent of the chemical composition. Indeed, the physical properties of the disk in X-ray models of a flared disk are largely a function of the vertical column density $N_{\perp}$, with $N_{\rm los}/N_{\perp}\approx10$. For a steady-state calculation in the absence of photodissociation (AGM11; NAG11), the molecular abundances of water and OH in the warm atmosphere are well approximated using $N_{\rm los}/N_{\perp}\!\approx30$. We use the same relationship in the calculations reported here and find that the results are insensitive to changes in this ratio by factors 
of as much as 3, which justifies using this approximation in place of a time-consuming full-disk analysis of the disk chemistry that includes FUV attenuation.  

In the absence of detailed information on the dust in the inner regions of protoplanetary disks, we use the dust model in Appendix A of GNI04, characterized by the dust to gas mass density ratio $\rhod / \rhog$ and the geometric mean grain size $\ag$. This model is based on the distribution of \citet{MRN} with minimum and maximum dust sizes, $a_1$ and $a_2$, and $\ag = \sqrt{a_1 a_2}$. The resulting grain surface area per H nucleus is
\be
\label{grain_surface}
S_{\rm d} = \frac{\nd \left<\pi a^2\right>}{\nh} = \frac{3}{4} \, 
\frac{\rhod / \rhog}{(\tilde{\rho}/1.35\mh)\ag},
\ee   
where $\tilde{\rho} \approx 3 \gpcc$ is the mean internal mass density of a grain, $\nH$ is the volumetric density of hydrogen nuclei, $\nd$ is the volumetric density of dust particles, and  1.35\,\mh\ is the mass per H nucleus. With these parameters and $\rhod / \rhog = 0.01$ and $\ag = 0.707~\micron$,  $S_{\rm d}=8\times10^{-23}$~cm$^2$. 
By choosing $\ag$ larger than the interstellar value of $\ag = 0.035 \micron$, the model takes into account that large grains settle to the midplane and a reduced population of small grains 
remains in the upper atmosphere.
The values used here and in past publications, $\rhod / \rhog = 0.01$ and $\ag = 0.707\,\micron$ or 7.07\,\micron, correspond to reductions in grain surface area of roughly ~20 and ~200 relative to the interstellar medium (ISM). 
We use the temperatures from the 
dust-gas mixture in \citet{DAlessio1999} as the dust temperature in all of our calculations. 

When we express the optical properties of protoplanetary dust in terms of standard absorption, scattering and extinction coefficients, $Q_{\rm abs}$, $Q_{\rm sc}$ and $Q_{\rm ext}=Q_{\rm abs}+Q_{\rm sc}$, the extinction cross section per H nucleus is
$\sigma_{\rm ext}=Q_{\rm ext}\,S_{\rm d}$ and the dust extinction associated with a hydrogen column $\Nh$ is
\be
\label{dustopdepth}
\taud = \sigma_{\rm ext} \Nh  = Q_{\rm ext} \, \Nh S_{\rm d}.
\ee
The grain surface area per H nucleus also plays a central role in the formation of \hm\ and in the thermal accommodation of gas and dust, as discussed in the next section. 

\subsection{Heating and Cooling}

\begin{figure}[h]\begin{center}
\includegraphics[width=3.25in]{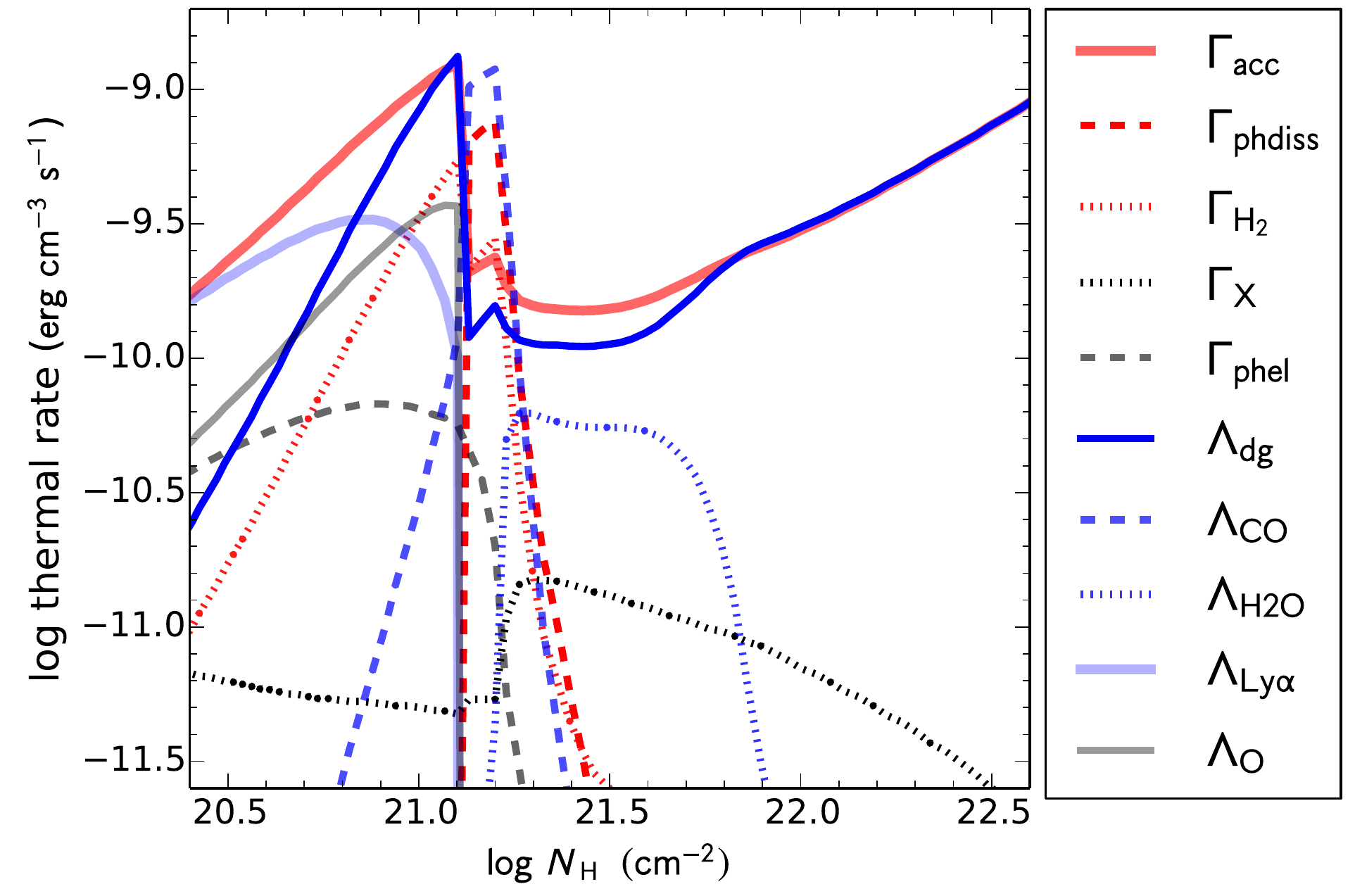}
\caption{\label{fig:heating_rates} Dominant thermal rates for heating $\Gamma$ ({\em red}) and cooling $\Lambda$ ({\em blue}) in the reference model of the disk atmosphere at $r=0.95$\,AU. The net gas cooling from thermal accommodation generally balances accretion heating. Minor heating and cooling rates, e.g., FUV heating by the photoelectric effect, are shown in gray. $\Gamma_{\mathrm{phdiss}}$ is the sum of \water\ and OH photodissociation heating.}
\end{center} \end{figure}

We revise and extend the treatment of the heating and cooling of protoplanetary disk atmospheres in GNI04, especially the heating mechanisms. We emphasize the heating mechanisms because the underlying physics of these processes is generally less well founded than the cooling. These include the heating associated with FUV radiation and the formation of molecular hydrogen on grains. Figure~\ref{fig:heating_rates} illustrates the roles of the most important heating and cooling processes for our reference model (Table~\ref{t:std}) at a radial distance of 0.95\,AU.
Dust-gas cooling is the most important coolant, whereas accretion heating dominates the heating. Other heating processes such as photodissociation and grain formation play roles over limited ranges of column density.
Water cooling has been included by adapting the work of \citet{NK93}. 
Other cooling mechanisms in the model, e.g., O~{\sc{i}} fine-structure lines and recombination lines, have rates that fall below the plotted range in Figure~\ref{fig:heating_rates}.

\begin{deluxetable}{lcl}[]
\tablecaption{\label{t:std}Reference Model Parameters}
\tablehead{
Parameter & Symbol & Value}  
\startdata 
Stellar mass           & $M_*$    &  0.5 $\MSun$   \\
Stellar radius         & $R_*$    &  2 $\RSun$     \\
Stellar temperature    & $T_*$    &  4000\,K       \\
Disk mass              & $M_D$    &  0.005 $\MSun$ \\
Disk accretion rate    & $\dot M$ &  10$^{-8}\Msun\,\pyr$\\
Dust to gas ratio      & $\rhod / \rhog$  & 0.01 \\
Dust grain size        & $\ag$    &  0.707 \micron   \\
Dust extinction        & $Q_{\rm ext}$ & 1.0 \\
X-ray luminosity       & $\Lx$    &  2 $\times 10^{30}\, \ergps$ \\
X-ray temperature      & $\Tx$    &  1 keV         \\
FUV luminosity         & $\LFUV$  &  5 $\times 10^{31}\, \ergps$ \\
Accretion heating      & $\ah $   &  0.5                 
\enddata
\end{deluxetable}

\break

\subsubsection{Accretion Heating} 

Viscous accretion heating is the most important heating process in our model. Without it, the surface layers would not be warm enough to take advantage of the warm neutral chemistry that is at the heart of our thermal-chemical model of the inner disk atmosphere. We continue to use the formulation in Eq.~12 of GNI04, 
\be
\label{accheat}
\Gamma_{\rm acc} = \frac{9}{4} \alpha_h \rho c^2 \Omega.
\ee
where $\Omega$ is the angular rotation speed and where, to a good approximation, the local mass density of the gas, $\rho \approx (1+4\xHe) \mH$, includes He but no heavier atoms. The helium abundance, $\xHe$, as all abundances in this paper, is defined relative to total hydrogen.  The isothermal sound speed for gas temperature \Tg\ is $c^2 = k\Tg/m$, with the mean particle mass given by
\be
m  = \frac{1+ 4\xHe}{1-\xHH +\xHe + \xel}\mH. 
\ee
This quantity varies substantially in the transition region where hydrogen changes from atomic to molecular form in the disk atmosphere. With these definitions, Eq.~\ref{accheat} can be written in a numerically useful form, with units of \erg\,\pcc\,\ps, as
\begin{align}
\Gamma_{\rm acc} =\ &  6.18\times10^{-23}\, \alpha_h \, 
(1-\xHH +\xHe + \xel) \nonumber \\
&\times \nh \, \Tg 
\left( \frac{M}{\MSun} \right)^{1/2} 
\left( \frac{\mathrm{AU}}{r} \right)^{3/2}.
\end{align}
Equation~\ref{accheat} is based on the hydrodynamic theory of a viscous accretion disk where the viscosity $\nu$ is expressed as $\alpha_h\rho c^2/\Omega$ \citep{SS73}. GNI04 regarded this as a phenomenological formula with $\alpha_h$ a parameter that could vary with position and would differ from the more familiar ``alpha'' parameter that describes the average accretion rate. Some support for this idea comes from simulations of the magnetorotational instability (MRI), e.g., \citet{HiroseTurner2011}, but a more complete understanding of dissipation in the MRI would help put disk accretion heating on a sounder footing. For the reference model in this paper, we take $\alpha_h = 0.5$.

\subsubsection{X-ray Heating} 

The heating associated with X-ray irradiation has essentially the same form as in GNI04,
\be
\label{xrayheat}
\Gamma_{\rm X} = Q \,\zeta\, \nH , 
\ee
where $Q$ is the heating per ion pair produced by X-rays and $\zeta$ is the X-ray ionization rate per H nucleus. 
Our treatment of the processes described in Sec.~2.2 of GNI04 has been substantially revised, with the result that $Q \approx 18$\,eV in protoplanetary disks \citep{GGP12}. We calculate the ionization rate according to AGM11. The X-ray luminosity, $\Lx = 2 \times 10^{30}$ erg\,s$^{-1}$ is the same as that 
adopted by GNI04 (Sec.~2.2), who chose the average YSO X-ray luminosity found with {\it Chandra} for stellar masses in the $0.8-1.2\,\Msun$ range for the Orion Nebula Cloud \citep{Garmire2000}. Our choice of \Lx\ is also close to the median value 
of the pre-main sequence	sample studied by \citet{Gudel2010}.

\subsubsection{Photodissociation Heating} 

Photodissociation of \water\ and OH produces fast H atoms that can heat the gas, 
as noted by BB09. The kinetic energy of the heavy products OH and O can also contribute, but this is a relatively small effect ($\sim$5\%). The photodissociation products can also be left in excited states. At sufficiently high densities, realized in some parts of protoplanetary disks, they may be  collisionally de-excited and lead to heating. In the case of \water, photodissociation leads to highly excited levels of the ground electronic state of OH. The fluorescence of these levels has been detected in protoplanetary disks in the mid-infrared with {\it Spitzer} and ascribed to photodissociation \citep[e.g.,][]{Najita2010, CN11}.
Calculations of the heating in \water\ photodissociation 
\citep{Crovisier1989, RodgersCharnley2005}
are consistent with roughly equal amounts of direct heating and fluorescence.
  
We estimate photodissociation heating by first calculating the difference between the photon and dissociation energies, 
\be
E_{\rm phdiss} = (h \nu - D_{\rm diss}),
\ee
for each molecule and each band. The dissociation energies $D_{\rm diss}$ are 5.13\,eV for \water\ and 4.41\,eV for OH. After averaging over the FUV bands, 
$\left<E_{\rm phdiss}(\water)\right> = 3.52$\,eV and 
$\left<E_{\rm phdiss}(\OH)\right> = 5.54$\,eV. These quantities enter the thermal balance through the heating rates per unit volume for molecule $i$ (\water\ or OH), 
\be
\Gamma_{\rm phdiss}(i) = G(i) \, n(i) \, \eta(i)
\left< E_{\rm phdiss}(i)\right>,
\ee
where $\eta$(i) is the fraction of $E_{\rm phdiss}$ going into heat. For \water\ this fraction is about 0.5 ($\sim 1.8$\,eV). A similar fraction might well apply to the photodissociation OH because the outcomes suggested by the cross sections calculated by \citep{vDD84}
indicate the production of significant fluorescence, e.g., the 6300\,\AA\ lines from the $^1D_2$ level of atomic oxygen. Photodissociation heating of \water\ and OH is weak unless one or both of these molecules is near maximum abundance, $\sim 10^{-4}$. For example, photodissociation heating becomes dominant in the narrow transition region from 
$\log \Nh=$ 21.1 to 21.2 in Figure~{\ref{fig:heating_rates}}.

\subsubsection{Photoelectric Heating} 

Another component of the heating from FUV radiation arises from the photoelectric effect on grains and PAHs. The underlying physics is complicated because it depends on the poorly known properties of the grains, such as the size and charge distributions. In addition to the grain surface area and the FUV flux, some of the relevant properties are the absorption coefficient \Qabs, the probability for ejecting an electron $Y$, and the energy of the liberated photoelectron, $E\ph$. All of these quantities depend on the incident photon energy. In view of the uncertainties in the physical and chemical properties of the grains and small particles, we adopt average values for most of the quantities that enter into the heating rate per unit volume,
\be
\label{phelheat}
\Gamma\phel  = Y E\ph \,  \Qabs \Sd \, \nH \,\FFUV(\lambda),
\ee
and use Eq.~\ref{grain_surface} for \Sd\ and set \Qabs=1. For the other grain properties we rely on the theory of \citet{WD01} developed for the ISM to estimate the first two factors in Eq.~\ref{phelheat}. The effective band for the photoelectric effect extends from $\sim$8\,eV to 13.6\,eV (911.7 -- 1550\,\AA), and the typical threshold (work function) is $\sim$5.5\,eV. On this basis, the mean photoelectron energy is $E\ph$ = 5.3\,eV.  \citet{WD01} give the optical properties of single silicate or carbonaceous grains as a function of photon energy and grain size. We estimate the efficiency to be $Y\!=\!0.05$ by roughly interpolating between a grain size of 0.03$\,\mu$m and bulk material in Figure~5 of \citet{WD01}. 
The efficiency of grain photoelectric heating will be reduced relative to the ISM with the reduction in the grain surface area  per H nucleus \Sd.\footnote{This is still an overestimate because we have ignored grain charging, which reduces grain heating according to the characteristic parameter, $(G/G_0)T^{1/2}/\nel$, where $G/G_0$ measures the ratio of the FUV radiation field to the standard value for the ISM. Using $G/G_0=10^6$, $T=900$K and $\nel \approx 3 \times 10^4\pcc$ for the upper active layer of the disk leads to $(G/G_0)T^{1/2}/\nel \approx 10^3$. When this value is used in Figure~16 of \citet{WD01} for ISM grains, reduction factors in the range 3 -- 5 are obtained.} 

\subsubsection{H$_2$ Formation Heating} 

Ever since the suggestion by \citet{Spitzer73} that newly formed H$_2$ molecules might carry a significant fraction of the dissociation energy (4.48\,eV), the amount of the resulting gas heating due to formation on grains has remained uncertain. The basic issue is how much energy goes into kinetic and internal excitation energy of the H$_2$ molecule in addition to the collective motions in the grain. Laboratory experiments \citep[e.g.,][]{Roser2003} suggest that the $\hm$ molecule thermally accommodates to the grain before making its final escape, at least as far as the kinetic energy is concerned. In a recent experiment, \citet{Lemaire2010} present evidence that $\sim$30\% of newly formed $\hm$ molecules are vibrationally excited for dust temperatures up to 70\,K. Assuming that the results of \citet{Lemaire2010} apply to protoplanetary disks, roughly 1.5\,eV in internal excitation of $\hm$ may be converted to heating by collisional de-excitation. Molecular dynamics calculations of $\hm$ formation via chemisorption on graphene surfaces suggest even larger values \citep{Sizun2010}. We assume here that the heating per newly formed  H$_2$ molecule is $\Delta E\approx1.5$eV.

We express the H$_2$ formation rate per unit volume in standard form as half the destruction rate of atomic H \citep{Tielens2010},
\be
\label{H2form}
R = \frac{1}{2} \bar v({\rm H}) n({\rm H}) \, \epsilon S(\Tg,\Td)\, \nH \, S_d ,
\ee
where $\bar v({\rm H}) n({\rm H})$ is the incident H atom flux, $\Sd$ is the grain surface area from Eq.~\ref{grain_surface}, $\epsilon$ is the formation efficiency and $S(\Tg,\Td)$ is the sticking probability that depends on both the gas and dust temperatures, \Tg\ and \Td.

Using the dust properties described above, we write the formation heating rate per unit volume (with units of \erg\,\pcc\,\ps) as,
\begin{align} 
\label{H2formheat}
\Gamma_{\rm H_2\, form} =\  6.91 & \times10^{-31} \, \Tg^{1/2} \, n({\rm H}) \, \nH \nonumber \\
& \times \epsilon S \,
\left( \frac{\rho_{\rm d}/\rho_{\rm g}}{0.01} \right)
\left( \frac{\mu m}{a_{\rm g}} \right)
\left( \frac{\Delta E}{{\mathrm{eV}}} \right).
\end{align}
In practice we follow GMN09 and use a revised estimate of curves 2 and 3 in the left panel of Figure~1 of \citet{CT2010} for the formation efficiency $\epsilon$ as a function of $\Td$. We approximate $\epsilon$=1 for \Td $<$ 25\,K, $\epsilon$=0.6 for $25 \le \Td <$ 80\,K, $\epsilon$=0.33 for $80 \le \Td <$900\,K, and 
$\epsilon$=0 for $\Td \ge$ 900\,K. We estimate $S=0.2$ for silicate grains from \citet{BH83}.

\subsubsection{Thermal Accommodation} 

When gas collides with dust particles, energy is exchanged in a classical kinetic theory process known as thermal accommodation. In principle the rate of energy exchange  depends on the nature of the gas and the dust surface, on the gas and dust temperatures,  and possibly on the dust particle size. We use the formulation in GNI04, rewriting their  Eq.~\ref{accheat} for the rate of energy exchange between gas and dust as,
\begin{align}
\label{dustgascool}
\Lambda_{\rm dg} =\ 4.76 & \times10^{-33}
{\cal{A}_{\rm{H}}} \sqrt{\Tg}\left(\Tg-\Td\right) \nH^2 \nonumber\\
& \times
\left(\frac{\rhod / \rhog}{0.01}\right)
\left(\frac{0.05\,\micron}{\ag}\right).
\end{align}
Temperatures are in K and densities in cm$^{-3}$ so that the units are erg cm$^{-3}$ s$^{-1}$. $\Lambda_{\rm dg}$ is defined so that it is a {\it cooling} function for $\Tg > \Td$, which is usually the case in the present calculations. The strength of the dust-gas cooling is determined by a mean accommodation coefficient $\cal{A}_{\rm{H}}$, which takes the form, omitting the small contributions of He and heavy atoms,
\be
\label{accommodation}
\mathcal{A}_{\rm H} = \xh \mathcal{A}(\h) + \frac{1}{\sqrt2} \xhm 
\mathcal{A}(\hm). 
\ee
The treatment given here of dust-gas cooling is based on the widely cited work of \citet{BH83}. These authors used a rather simple semi-classical theory to fit experiments on tungsten. They then calculated accommodation coefficients for H and H$_2$ on water ice, silicates and graphite. We fit  Figure~4 of \citet{BH83} for the individual coefficients, and in the present application assume that the dust is pure silicate. Laboratory measurements of the accommodation coefficients for astrophysical dust analogs for temperatures relevant to the inner regions of protoplanetary disks, $\Td \approx 100$\,K and $\Tg \approx 300-1000$\,K, would be very helpful.

\section{Results}

The results presented here for the FUV and X-ray model of the previous section are obtained by integrating the time dependent heat equation for the gas temperature $\Tg$ and the chemical rate equations for 117 species, using the \citet{DAlessio1999} flared disk model, which specifies the disk structure in terms of $\nh$ and $\Td$. The model is divided into cells located by their radial and vertical coordinates, $r$ and $z$. At each radius $r$, we begin at the ``top'' of the atmosphere, specified by the vertical column density 
$\log\,\Nh$=18, where we assume atomic conditions and optically thin dust. The numerical integrator \texttt{LSODA} is invoked via the \texttt{scipy.integrate} library\footnote{http://www.scipy.org} and run until steady state is achieved. Grid steps toward the mid-plane are evaluated using the steady-state abundances from overlying grid cells for the initial conditions of subsequent integrations. The chemical equations contain 1236 reactions, which are basically the same as in AGM11 and NAG11, except for minor improvements and the photodissociation reactions (\ref{R:phH2O}, \ref{R:phOH}) described in Sec.~2.1. Table~\ref{t:std} gives the most important model parameters for our reference model.

\subsection{The Molecular Transition}

\begin{figure}[h]\begin{center}
\includegraphics[width=3.25in]{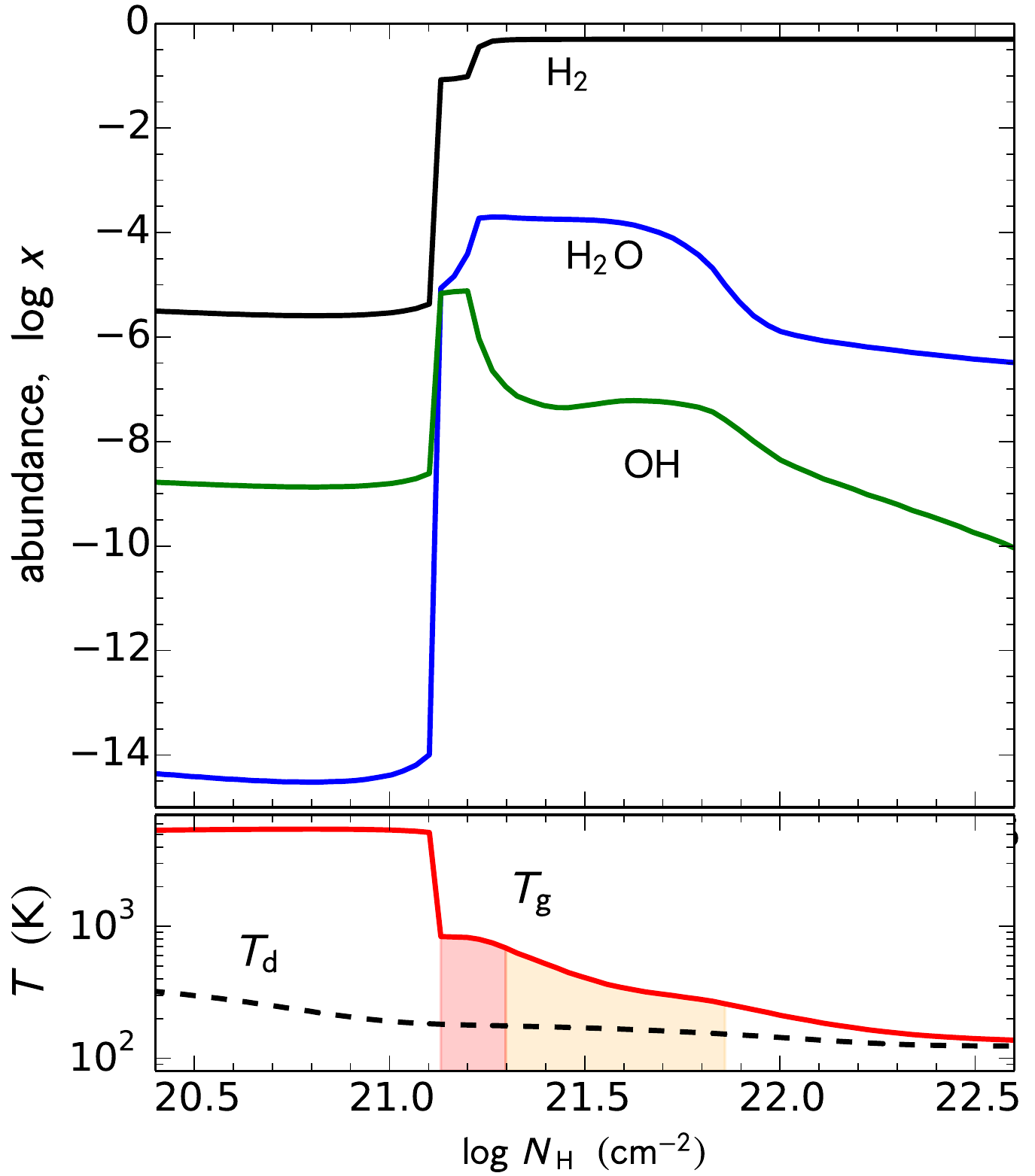}
\caption{\label{fig:standard}Vertical profiles of molecular abundances ({\em top}) and temperatures ({\em bottom}) in the reference model of the disk atmosphere at $r=0.95$\,AU. The red shaded region is where $\Tg$ is in the range 650 -- 1000\,K, and the orange shaded region is where $\Tg$ is in the range 250 -- 650\,K.}
\end{center} \end{figure}

Figure~\ref{fig:standard} gives the run of chemical abundances vs.~depth through the inner disk atmosphere for the reference model at $r$=0.95\,AU. The abundances are defined in terms of volumetric densities relative to $\nh$. The height is expressed as the vertical column density of total hydrogen nuclei, \Nh, measured from the top of the atmosphere. The abundance of all of the major species of interest, H$_2$, OH, \water\ and CO (not shown), increase rapidly when the vertical column $\Nh$ approaches $\log\Nh=21.1$, delineating the transition to the molecular region. At this point CO takes up essentially all of the carbon, H$_2$ has 20\% of the hydrogen, and the abundances of both \water\ and OH are in the $10^{-6}-10^{-5}$ range. At somewhat larger columns $\log \Nh \approx 21.3$, essentially all of the hydrogen is in H$_2$, and \water\ has taken up all of available oxygen not in CO.

These transitions start at the location in the hot atmosphere where the temperature begins to decrease from a characteristic level of $\Tg \approx 5000$\,K (in this case at $\log\Nh=21.0$). Referring to Figure~\ref{fig:heating_rates}, this occurs where dust-gas cooling begins to dominate Ly$\alpha$ cooling with increasing density and to balance accretion heating. Concomitant with the formation of molecules, CO cooling rapidly forces the temperature down close to $\sim$1000\,K.
The transition region includes a column of $\sim$10$^{19}$\,\psqcm\ of H$_2$ in the temperature range \Tg=1000 -- 4000\,K in the inner disk. These conditions are similar to those inferred from {\it HST} observations of H$_2$ fluorescence \citep{Herczeg2004, Schindhelm2012, France2011, France2012a}.

The temperature of the transition region in Figure~\ref{fig:standard} falls slowly below 1000\,K, reaching $\Tg = 250$\,K for a vertical column $\log\Nh = 21.9$. Thus the transition region in the X-ray and FUV irradiation model consists of roughly three layers of gas with different temperatures and chemical properties: 
(1) a warm molecular region, which lies below 
(2) a warmer photodissociation region where H$_2$ and \water\ change from partial to full association, and 
(3) a hot atomic region at the top of the atmosphere.
Regions (1) and (2) are distinguished in Figure~\ref{fig:standard} by the light orange and red shading, respectively. 
Throughout the disk the temperature of $\sim$650\,K 
roughly separates the region of warm \water\ from warmer OH.
At the start of the transition, where the temperature is near 1000\,K, OH is as abundant as \water. By contrast, below the transition region, the temperature drops below 250\,K and the abundance of OH is typically 3-4 dex smaller than that of \water. Once the dust temperature falls below $\sim$125\,K freeze out of \water\ has to be taken into account. At the radius $r=0.95$\,AU in Figure~\ref{fig:standard}, freeze out would occur where $\log\Nh > 22.0$ if the freeze out temperature for \water\ is 125\,K.

\begin{figure}[h]\begin{center}
\includegraphics[width=3.5in]{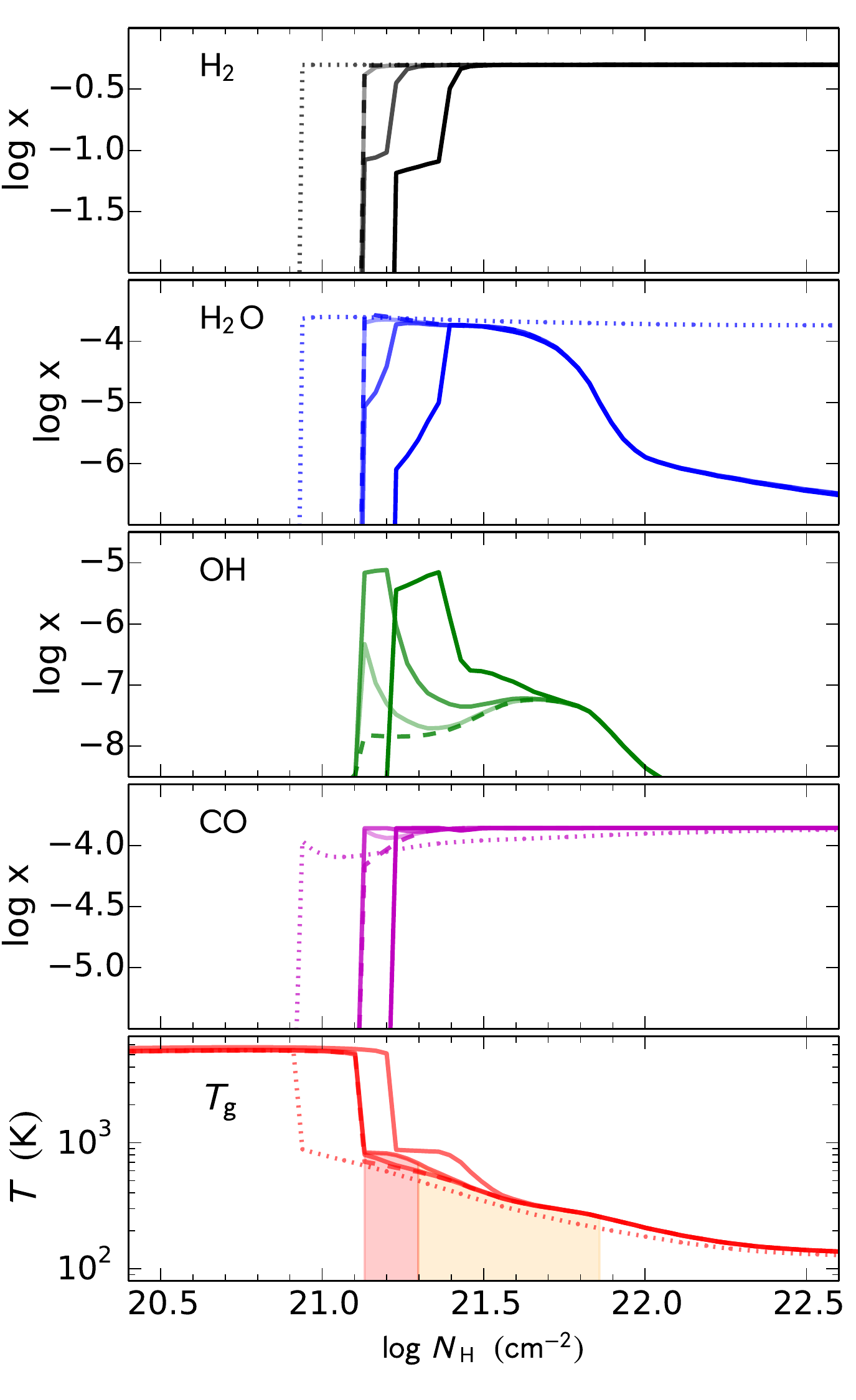}
\caption{\label{fig:roleFUV}Vertical profiles of molecular abundances for H$_2$ ({\it black}), \water\ ({\it blue}), OH ({\it green}), and CO ({\it magenta}), and gas temperatures ({\it red}) in the disk atmosphere at $r=0.95$\,AU. Dotted curves are abundances without irradiation. Dashed curves are abundances under X-ray irradiation alone. Solid lines correspond to $\LFUV$=0.0013, 0.013, and 0.13\Lsun, from faint to dark respectively. $\Lx=5.2\times10^{-4}\LSun$ in each model with X-rays. For H$_2$ and \water, the 0.0013\Lsun\ case essentially overlaps X-ray irradiation alone. The red shaded region is where $\Tg$ is in the range 650 -- 1000\,K, and the orange shaded region is where $\Tg$ is in the range 250 -- 650\,K.}
\end{center} \end{figure}

Further insight into the nature of the transition region is provided in Figure~\ref{fig:roleFUV} where the effects of varying the X-ray and FUV luminosity are illustrated. The abundance curves at $r=0.95$\,AU are displayed for models with neither X-ray nor FUV radiation (dotted curves), X-rays in the absence of FUV (dashed curves), and variations of FUV that range from 0.1 to 10 times the reference FUV luminosity value, \LFUV= 1.3 $\times10^{-2}\Lsun$ (the solid curves) at fixed X-ray luminosity. 
The atomic to molecular transition in the inner region of an FUV and X-ray irradiated protoplanetary disk occurs in steps over a layer of finite thickness. 
This is to be contrasted with abundances calculated in a model with neither FUV nor X-ray irradiation.
Indeed, in the absence of X-ray and FUV irradiation, the abundances of H$_2$, \water\ and CO all rise up sharply at $\log\Nh = 20.9$ and contain essentially all of the available hydrogen, oxygen, and carbon. Inclusion of X-rays without FUV shifts the transition down to $\log\Nh = 21.1$. However, the atomic to molecular transition still occurs in a very thin layer. When both X-rays and FUV are included, the molecular abundances in Figure~\ref{fig:roleFUV} rise rapidly at $\log\Nh = 21.1$, close to the pure X-ray case, but not to their maximum values. Beyond this depth, the H$_2$O abundance continues to increase, controlled by photodissociation mediated by absorption of the FUV by dust and then by self-shielding. Maximum abundances of H$_2$ and \water\ are achieved at $\log\Nh = 21.2$ in the reference model. 
One result of larger FUV is to increase the amount of OH and decrease the amount of \water\ in the warm photodissociation region. The gradual atomic to molecular transition is a special characteristic of a protoplanetary disk that is irradiated by both X-rays and FUV. 

\subsection{The Chemistry of Water and OH}

The chemistry is governed by the familiar temperature-sensitive radical reactions used by GNI04 and in subsequent papers, e.g, GMN09 and NAG11. The formation of OH plays a key role, starting with the endothermic reaction,
\be
\label{OHform}
{\rm O} + \hm \rightarrow {\rm H} + {\rm OH}; \tag{R3}
\ee 
the rate coefficient recommended by \citet{Baulch2005} is
\be
\label{neutreacOH}
k_3 = 8.5\times 10^{-20}\,T^{2.67}\, e^{-3163/T} {\rm cm}^3 \ps.
\ee
Then water is made by an exothermic reaction with a modest barrier,
\be
\label{waterform}
{\rm OH} + {\rm H}_2 \rightarrow \water + {\rm H}; \tag{R4}
\ee 
its rate coefficient is
\be
\label{neutreacH2O}
k_4 = 8.5\times 1.70 \times 10^{-16}\,T^{1.60}\, e^{-1660/T} {\rm cm}^3 \ps.
\ee 

These reactions are effective for intermediate temperatures ranging from $\sim$250\,--\,1000\,K.  Below 250\,K, the barrier in Eq.~\ref{neutreacOH} suppresses formation, whereas above 1000\,K the barrier to the inverse reaction in \ref{OHform} can be overcome. 
For example, in the reference model at the bottom of the hot layer just below 
$\log\Nh = 21.0$, the balance of forward and backward reactions in \ref{OHform} and \ref{waterform} leads to OH abundances in the range $10^{-8} - 10^{-6}$ and to much smaller but rapidly growing \water\ abundances. Once the temperature drops to $\sim$1000\,K, the inverse reactions are significantly reduced and the dominant destruction pathway for OH and \water\ becomes FUV photodissociation.

The water abundance increases with increasing depth and density until its maximum value is reached just after $\log\Nh = 21.2$. At this point, consumption of H$_2$ by the above reactions is reduced, H$_2$ reaches its maximum abundance, and OH decreases as almost all oxygen is incorporated into \water. In the layer where the \water\ abundance increases from $10^{-6}$ to $2 \times 10^{-4}$, FUV photodissociation is the dominant, although not the only, destruction mechanism. 

The increase in \water\ is moderated by the absorption of FUV, by dust and by \water\ itself, which is the self-shielding process discussed in BB09. 
In the FUV dissociation region of the reference model in Figure~\ref{fig:standard}, $\taud > \tau_{\water} + \tau_{\OH}$ midway to where the water 
abundance 
peaks ($\log\Nh \approx 21.1$), at which point the total FUV optical depth is large. 
Once \water\ reaches its maximum abundance, destruction by reactions with  X-ray generated molecular ions become more important than photodissociation, although one of the major ions, 
H$_3$O$^+$, is partially recycled to \water\ by dissociative recombination. Going deeper down into the atmosphere, where $\log\Nh > 21.8$ at $r=0.95$\,AU, the temperature continues to drop and OH is no longer rapidly converted into \water\, according to the rate coefficient in Equation~\ref{neutreacH2O}, and instead fast exothermic reactions with atomic oxygen lead to high abundances of O$_2$. The abundance of OH is also reduced, so that  most of the oxygen is in the form of O and O$_2$. The transition from  \water\ to O$_2$ and O occurs for all radii larger than 0.5\,AU.

\subsection{The Distribution of Water and OH}

\begin{figure*}[]\begin{center}
\includegraphics[width=6.75in]{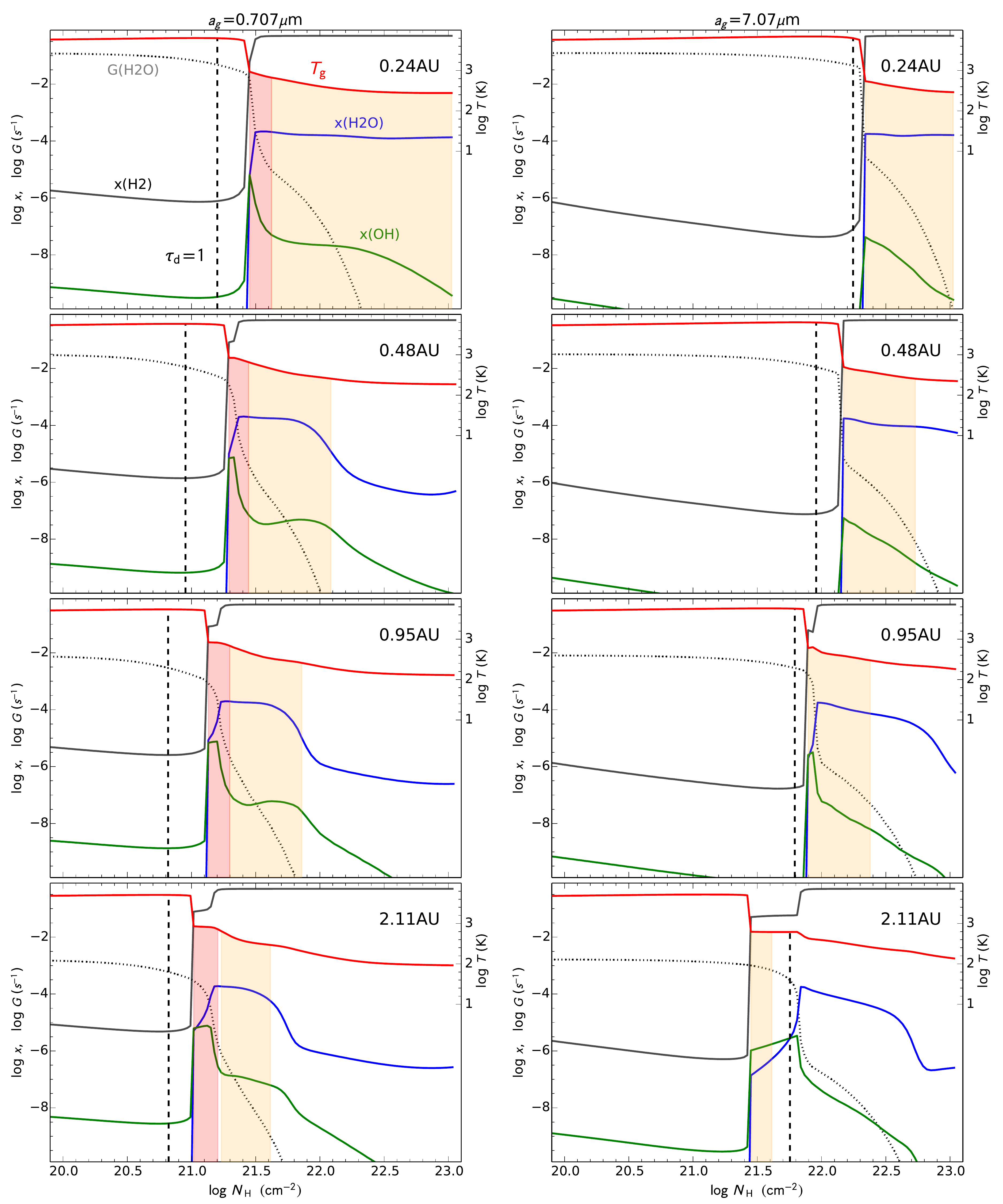}
\caption{\label{fig:warm_regions}Vertical distributions of abundances for H$_2$ ({\em black}), \water\ ({\em blue}) and OH ({\em green}) 
for the reference model ({\em left}) and 
the reduced grain area case ({\em right}) at several radii, 
i.e., with $\Lx=5.2\times10^{-4}\LSun$ and 
$\LFUV=1.3\times10^{-2}\LSun$. 
$\Tg$ is plotted ({\em red}) with units along the right axis. The red shaded region is where $\Tg$ is in the range 650 -- 1000\,K, and the orange shaded region is where $\Tg$ is in the range 250 -- 650\,K. 
The line of sight dust FUV optical depth of unity is indicated by the vertical ({\em dashed black}) lines. Curves for the dissociation rate of water, $G$, are shown ({\em dotted black}). 
}
\end{center} \end{figure*}

Figure~\ref{fig:warm_regions} shows how the abundances of \water\ and OH in the disk atmosphere change with radius and grain surface area, with the reference case ($\ag = 0.707\mu$m) on the left and the case where the grain surface area is reduced by another factor of 10 ($\ag = 7.07\mu$m) on the right. 
Results are shown only for small radii, $r \lesssim 2$\,AU, where freeze out does not occur until below the regions of maximum gaseous abundance of \water\ and OH. 
The photodissociation rates for \water\ are plotted as dotted curves. 
The unshielded rate at $r=0.95$\,AU is $G$=10$^{-2}\,\ps$, which is 10$^7$ times larger than the value for water in the ISM.
The effects of self-shielding are evidenced by the sudden drop in $G$ near where the water abundance peaks. These curves for $G$ apply over the entire range of vertical columns in Figure~\ref{fig:warm_regions}, even where water is subject to freeze out. In such regions water vapor may still be produced by the warm chemistry in reactions \ref{OHform} and \ref{waterform} because the progenitor species (H$_2$, O and OH) do not freeze out until much lower temperatures. In order for water to freeze out, adsorption onto grain surfaces has to successfully compete with photodissociation.

In the reference case on the left side of Figure~\ref{fig:warm_regions}, the thickness of the warm
(shaded orange) region decreases with increasing radius, while the peak abundances remain high. The thickness of the warm region mainly affects \water\ because its maximum abundances are achieved at greater depths than for OH. Beyond $\sim$0.5\,AU, the \water\ abundance always decreases when the temperature falls below $\sim$300\,K. The variation of the \water\ and OH abundances with radius shown in Figure~\ref{fig:warm_regions} 
is another manifestation of the role of temperature in the warm radical chemistry. 

The panels on the right side of Figure~\ref{fig:warm_regions} lack 
the red shaded regions, where $\Tg \approx 650 - 1000$\,K, illustrating that the amounts of warmer OH and \water\ are significantly reduced by decreasing the grain surface area.
When the grain area is reduced, the gas is not thermalized as efficiently 
and the atmosphere remains hot to larger \Nh, causing the molecular transition to occur deeper, where densities are larger and the dust is colder. At $r=2$\,AU, near the top of the transition, the small grain area model
has larger warm abundances of OH than \water. This is a direct effect of the role of dust in providing less attenuation of the FUV, as indicated by the plots of the photodissociation rates. The dashed vertical lines in Figure~\ref{fig:warm_regions} indicate the location where the FUV dust optical depth, from Eq.~\ref{dustopdepth}, is unity. Increasing the dust grain size reduces the dust surface area $\Sd$ according to Eq.~\ref{grain_surface} and lowers the FUV opacity. The value $\taud = 1$ occurs near $\log\, \Nh$ = 20.8 at $r=2$\,AU in the reference model, before water self-shielding becomes important. In the case of the reduced grain surface area at $r=2$\,AU, the FUV penetrates much deeper into the disk, beyond the transition region, and the large photodissociation rates throughout reduce the abundance of \water\ in the warm region, producing OH, and fully self-shielding at $\log \Nh$=21.8.

In order to compare with observations, we first calculate the vertical column densities of warm \water\ and OH as a function of radius and then integrate them to obtain the cumulative number of molecules within a given radius. The vertical column densities are given in Table~\ref{t:columns} for the two temperature regimes, 250 -- 650\,K and 650 -- 1000\,K, which correspond roughly to where \water\ and OH, respectively, are most abundant. Results from both the 
reference grain area ($a_g = 0.707 \mu$m) and the reduced grain area ($a_g = 7.07 \mu$m) cases 
are tabulated. To be conservative, freeze out of \water\ onto grains has been taken into account by excluding gaseous water from regions where the dust temperature is less than 125\,K. 

\begin{deluxetable}{ccccc}[]
\tablecaption{\label{t:columns}Column densities of H$_2$O and OH}
\tablehead{
$r$ & 
\multicolumn{2}{c}{$\log N(\water)$} & \multicolumn{2}{c}{$\log N({\rm OH})$}   \\ 
(AU)    &  250-650\,K & 650-1000\,K &  250-650\,K & 650-1000\,K}  \\
\startdata
\multicolumn{5}{c}{Reference model, $\ag=0.707\mu$m }\\ 
\tableline
  0.24   &   19.1    &   17.4    &   14.8    &   15.3     \\ 
  0.48   &   18.0    &   17.1    &   14.5    &   15.4     \\ 
  0.95   &   17.7    &   16.9    &   14.4    &   15.4     \\ 
  2.11   &   17.5    &   16.7    &   14.4    &   15.5     \\ 
  5.25   &   ....    &   14.8    &   ....    &   14.9     \\
\tableline
\multicolumn{5}{c}{
Reduced grain area, $\ag=7.07\mu$m } \\
\tableline
  0.24   &   19.1    &   ....    &   14.7    &   ....     \\ 
  0.48   &   18.6    &   ....    &   14.6    &   ....     \\ 
  0.95   &   18.2    &   ....    &   15.6    &   ....     \\ 
  2.11   &   14.6    &   ....    &   15.3    &   ....     \\ 
  5.25   &   ....    &   ....    &   ....    &   ....    
\enddata
\tablecomments{The units for column densities are cm$^{-2}$.}
\end{deluxetable}

\begin{figure}[h]\begin{center}
\includegraphics[width=3.25in]{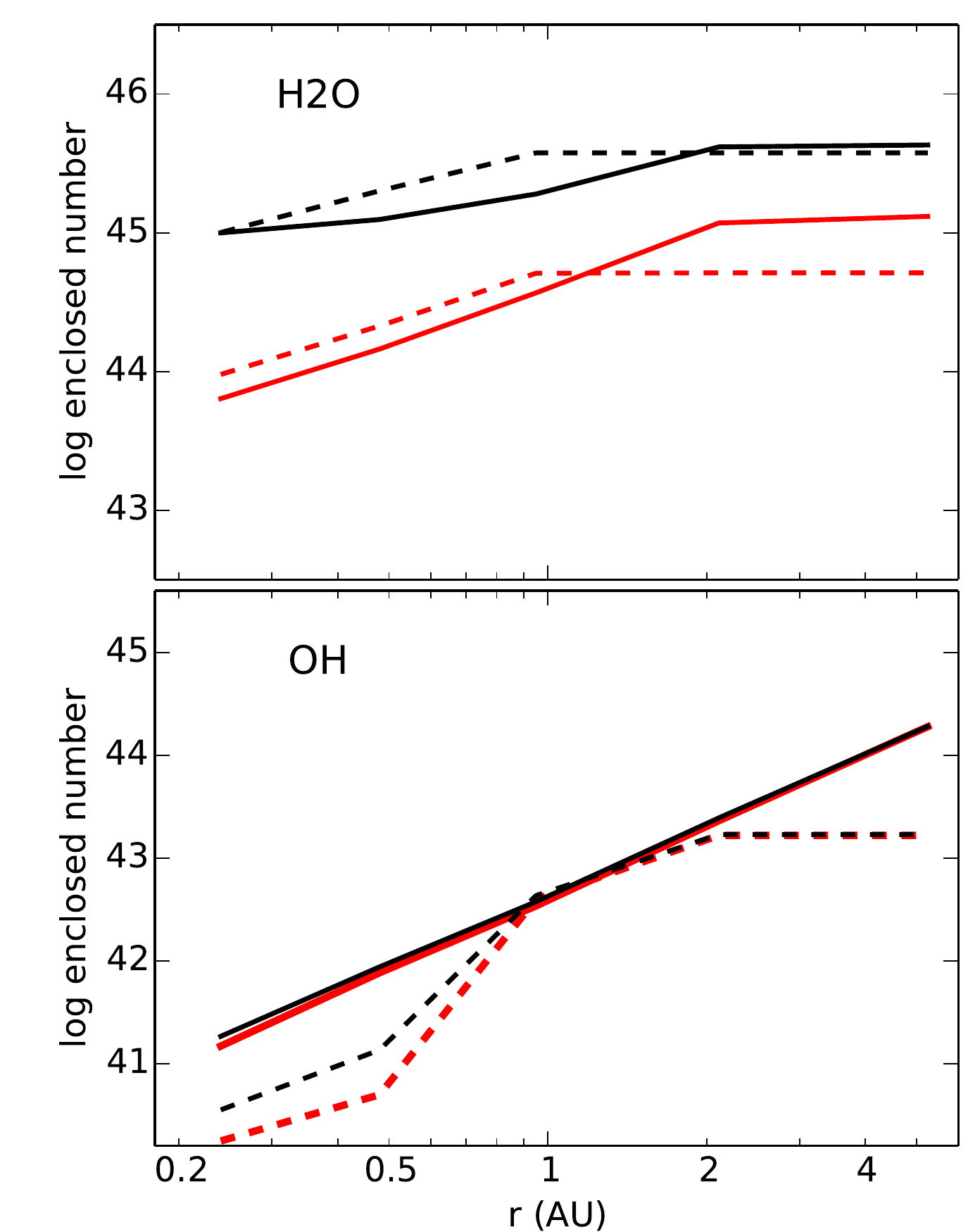}
\caption{\label{fig:enclosed_mol} The number of warm molecules within a radius $r$, with \Tg\ in 
the range 250 -- 1000\,K, and where $\Td > 125$\,K (i.e., above the freeze out temperature) are shown in {\em black} for both the reference model ($\ag=0.707\,\micron$, {\em solid lines}), and for 
reduced grain area 
($\ag=7.07\,\micron$, {\em dashed lines}). Particularly
warm molecules with \Tg\ in the range 450 -- 1000\,K, are shown in {\em red}.}
\end{center} \end{figure}

The freeze-out temperatures for oxygen-bearing species O, O$_2$, OH and CO are all much lower and these species stay in the gas phase in the inner disk. Since OH is most abundant in warm regions where  \water\ remains in the gas phase, freeze-out in cold regions does not significantly impact our calculation of OH columns. Where \water\ freezes out, the back-reaction in \ref{waterform} is slow, hydrogen is entirely molecular, and the OH chemistry is dominated by reactions with molecular ions and O rather than the photodissociation of \water.
In the reference case for \water, the largest columns occur within 0.5\,AU for the temperature range 250 -- 650\,K. The relatively slow decrease for larger radii in the reference case is compensated by the increase of annular area with radius.  The total numbers of warm \water\ and OH enclosed within a given radius are shown in Figure~\ref{fig:enclosed_mol}. The case of 
reduced grain area is distinguished from the reference case by the use of dashed lines. The total amount of warm \water\ levels off beyond $\sim$4\,AU due to freeze out. The total number of warm water molecules in the reference case is $\sim4 \times 10^{45}$ or $1.22 \times 10^{23}$\,g, which is about 9\% of the mass of the Earth's oceans. 

\section{Discussion}

The atmosphere of a protoplanetary disk that includes both FUV and X-ray irradiation 
consists of an upper hot region with $\Tg \approx 5000$\,K, a cool lower region with temperatures $\Tg < 250$\,K, and in between a warm layer with temperatures $ 250 < \Tg < 1000$\,K. The transition from atoms to molecules occurs mainly in the warm layer where large amounts of water are produced. 
For our reference model, gaseous water would be present in the region where the grain temperature is above the freeze out temperature ($\sim$125K), i.e., within a radial distance of 4\,AU. 
However, processes that we did not consider here, such as photodesorption and gas-grain chemistry, could well maintain a gaseous water reservoir over a larger range of radii.

We confirm several of the ideas presented by BB09; for example, molecular shielding by \water\ and OH is important in accounting for the large amounts of warm \water\ and OH that are found in {\it Spitzer} observations of FUV-irradiated CTTS disks. In addition, we also find that when significant grain settling has occurred, UV absorption by \water\ and OH plays an important role in the thermal and chemical properties of the disk atmosphere, and it shields the disk midplane from UV radiation.

In confirming these ideas, we have made several important improvements in the implementation of molecular shielding. One example is the treatment of molecular hydrogen, which plays a key role in the formation of OH and \water. BB09 assumed that all hydrogen was in H$_2$ in the atmosphere without considering the chemistry of H$_2$, especially the role of grains.  They assumed molecules dominated the FUV opacity due to dust depletion, but did not examine whether depletion was consistent with the abundance of H$_2$. We find that molecular shielding is effective even in the presence of significant grain settling, i.e., at the level of grain settling seen in T Tauri stars. 

When we examined the model results in Sec.~3 to see how FUV radiation changes the pure X-ray model, we found that it is the X-rays that still play a role at the start of the transition. However, the rise up to maximum molecular abundances is achieved more gradually under the influence of FUV photodissociation of \water\ and OH. Once the FUV radiation has been largely extinguished, the X-rays again play a significant role by reshuffling oxygen between the molecular ion H$_3$O$^+$ and O, OH, and \water. However, it is the temperature structure that dominates the general character of the molecular transitions. With characteristic activation energy barriers measured in thousands of K, the rate coefficients for the reactions that produce water (\ref{OHform}, \ref{waterform}) can change by many orders of magnitude throughout the protoplanetary disk atmosphere.

Both the large and small grain area models produce similar amounts of water in the warm atmosphere (Fig~\ref{fig:enclosed_mol}, upper panel). Thus, water emission should be detectable even from disks that have experienced a large amount of grain settling. In either model, the number of warm water molecules ($\sim$10$^{45}$) is in agreement with quantities derived from {\it Spitzer} observations \citep{CN11, Salyk2011}. Despite the similar amounts of water produced in the two models, we find some potentially observable differences. A greater fraction of the water is warm (in the 450K -- 1000K range) in the model with the larger grain area. The water also arises over a larger range of radii in the larger grain area model and would produce more centrally peaked line profiles for the water emission than in the smaller grain area model. 

The water is mainly produced by the warm radical chemistry used in previous models, starting with GNI04 and in most other chemical models cited in Sec.~1. The pre-condition for the formation of oxygen-bearing molecules is the presence of H$_2$ (formed on dust grains), from which OH and \water\ are produced by reactions with O and OH, respectively. If the temperature is too low, thermal barriers reduce the formation rates, and if the temperature is too high, inverse reactions dominate formation. OH and \water\ are found in the inner regions of protoplanetary disks where the gas temperatures lead to efficient reactions of H$_2$ with the neutral radicals O and OH. In this way the abundances of OH and \water\ are intimately tied to the chemistry of H$_2$, as emphasized by GMN09. But the production of these molecules is also linked to the properties of the dust in protoplanetary disk atmospheres by several processes: the dust surface is essential for forming H$_2$, for cooling the gas by thermal accommodation, for extinguishing the incident FUV, and for heating the gas by the photoelectric effect. 

According to Figures~\ref{fig:standard} and \ref{fig:warm_regions}, abundant OH and \water\ are found at different places in the disk atmosphere: the OH is higher and hotter than the \water. Looking down into the disk atmosphere, a layer of the OH appears first and below it layers of warm and cool \water. A similar distinction in the OH and \water\ temperatures may be present in {\it Spitzer} observations of CTTS disks. \citet{CN11} analyzed {\it Spitzer} observations of the disks of six CTTSs in Taurus-Aurigae, and obtained temperatures of $\sim$600\,K for \water. Working with {\it Spitzer} observations of a larger and different sample, \citet{Salyk2011} found \water\ temperatures clustering around $\sim$450\,K and significantly higher OH temperatures, with an average value of $1050$\,K for 15 detections. 

While similar amounts of \water\ are produced in the large and small grain area models, greater differences are seen in the OH that is produced in the two models. As shown in Figure 6 (lower panel), the large grain area model (solid line) produces much more OH than then small grain area model (dashed line). The enclosed number of warm OH molecules in the large grain area model is similar to that observed by {\it Spitzer} ($\sim 10^{44}$).  In either model, the OH arises from a larger range of radii than the \water. Taken literally, the OH and \water\ line profiles would differ, with the OH profiles more centrally peaked than the \water.  

Our model does not treat the the \Lya\ line, which can be a significant fraction of the incident FUV. For example, \Lya\ accounts for $\sim$80\% of the observed total FUV for TW Hya \citep[BB09;][]{Yang2012}. The central portion of the observed line is absorbed out, and it appears as two broad features separated by several \AA. Following \citet{BB11}, the hot upper atmosphere of the disk is a region of diffuse \Lya\ due to multiple scattering of the line by atomic H. In our reference model discussed in Sec.~3, atomic H is the dominant form of hydrogen almost down to vertical columns of $\log \Nh \approx 21.2$. Thus the diffuse \Lya\ extends throughout the upper part of the warm atmosphere to the level where OH has begun to decrease from its maximum abundance and \water\ has achieved its maximum. 

\citet{BB11} used a Monte Carlo code to calculate the transfer of \Lya\ for a D'Alessio disk model that includes a simplified chemistry for the H$_2$ molecule and three levels of dust depletion. If the dust surface area is reduced by the value 20, as in our reference model, it can be deduced from their Figure~10 that the initial ratio of 6 for the \Lya\ to the FUV continuum number flux is reduced to $\sim$1/3 near $\log \Nh = 21$. On this basis, the \Lya\ line  would increase the dissociation rate of water beyond that calculated in this paper by $\sim$1/3. The increased photodissociation could be approximated by increasing \LFUV\ in our model, with the results illustrated graphically in Figure~\ref{fig:roleFUV}.

The role of the absorption of \Lya\ by \water, ignored by \citet{BB11}, is also important. Using the cross section for \Lya\ absorption by \water\ \citep[$1.6\,\times\,10^{-17} \sqcm$,][]{Mota2005}, the vertical optical depth through $\log \Nh = 21.0$ is $\sim$3.2, ignoring the fact that the \Lya\ photons are not all traveling downward. This reduces the effect of the \Lya\ line by a further factor of 25 beyond the effects of dust included by \citet{BB11}. On this basis we do not expect the total amount of \water\ estimated in this paper to be much affected by the omission of \Lya. The details of the transition region may change, as may the properties of the OH radical, which is located in a region with more diffuse \Lya\ radiation. Thus it would be of interest to include the transferof \Lya\ in our model of a thermally and chemically inhomogeneous protoplanetary disk atmosphere.

It is interesting to compare our relatively simple model with the very detailed model, ProDiMo. Presented by Woitke et al.~(2009), this model was extended and applied in several subsequent papers, especially Meijerink et al.~(2012), which includes X-ray as well as FUV irradiation. ProDiMo includes a very large number of thermal processes. For the purposes of producing the warm temperatures in the inner disk, ProDiMo relies on photoionization of PAHs and collisional de-excitation of excited Fe\,II levels. Our main heating mechanism stems from mechanical heating that can be traced to the dissipation produced by the accretion process, most likely the MRI. There are uncertainties associated with these processes, e.g., the PAH abundance in disks and the magnitude of mechanical heating. In both models, dust-gas cooling, discussed in Sec. 2.2.6, is important. 

Another difference between ProDiMo and our model is their use of a relatively large inner radius of 0.5\,AU; essentially all of the warm water predicted by Woitke et al.\ arises in the very inner rim of the disk rather than over the  inner several AU as in our model. Low \water\ abundances ($x_\water \lesssim 10^{-7}$) are found beyond the inner rim, in contrast with the much larger abundances in our model ($x_\water \approx 10^{-4}$). These differences imply that ProDiMo would have difficulty accounting for the large amounts of \water\ that have been observed in inner CTTS disks if it adopted a smaller inner radius more typical of CTTS. A likely reason for this major difference between the results from ProDiMo and our model is  that we have included UV shielding by OH and \water, which both absorb  radiation over a wide range of FUV wavelengths, whereas ProDiMo includes UV shielding by CO and H$_2$, which shield over a more limited range of wavelengths. 

\section{Conclusions}

We have developed an integrated thermal-chemical model for protoplanetary disk atmospheres that includes 
irradiation by FUV and X-rays, grain settling, and a detailed treatment of the physical processes that affect water and other molecules. Our objective has been to identify some of the key processes and how they operate. We emphasized that several of the processes are uncertain, especially those 
relating to grains, and we have focused on a reference model that represents a typical CTTS disk. 

Our results support the idea that \water\ in the inner regions of protoplanetary disks can be formed {\it in situ}. This thesis has been discussed in several of the earlier modeling papers cited in Section 1. Especially noteworthy are those that obtain sufficient warm columns of \water\ to be consistent with the amounts deduced from Spitzer observations, e.g., Glassgold et al.~(2009), Woods \& Willacy (2009), Heinzeller et al.~2011, Najita et al. (2011). In the present work, both the physical and astrophysical details of how adequate amounts of \water\ can be formed are analyzed in detail. Among the new results presented here is the key role of dust surface area in regulating processes that directly affect the \water\ abundance, i.e., H$_2$ formation, dust-gas thermal accommodation, and the extinction of FUV radiation.  All of these processes are sensitive to the size distribution of the grains in the upper atmosphere of the disk. We find that, in our updated model, large amounts of \water\ and OH are synthesized in the observable warm regions of a typical protoplanetary disk atmosphere even in the presence of FUV and X-ray radiation and substantial grain settling. 

The \water\ and OH are located primarily in distinct locations, with the warm \water\ layer located below a warmer layer of OH. Both the temperatures and total numbers of molecules are consistent with {\it Spitzer} observations of CTTS, with the OH in the atmosphere extending over a larger range of radii than the \water. In our model, freeze out on dust grains restricts the gaseous water reservoir in the warm atmosphere to radii $< 4$\,AU. 

UV absorption by \water\ and OH is important in accounting for the large amounts of warm water that are observed, confirming the ideas discussed by BB09. FUV radiation affects the width and sharpness of the atomic to molecular transition, whereas X-rays and grain settling affect the depth at which the transition starts. X-rays also generate ions that destroy water below the photodissociation region. Many of these conclusions can be tested with suitable observations, as has already been done in some cases. Together with the discussion in previous sections, they present a richer picture of an atomic to molecular in protoplanetary disk atmospheres. 

\vspace{2ex}
We acknowledge support from NASA grant NNG06GF88G (Origins) and NASA grant 1367693 
({\it Herschel} DIGIT). We are particularly grateful to Paula D'Alessio for providing 
the disk model used in this work, and we will remember her for her kindness and her many contributions to the understanding of protoplanetary disks.


\begin{thebibliography}{68}
\expandafter\ifx\csname natexlab\endcsname\relax\def\natexlab#1{#1}\fi

\bibitem[{{{\'A}d{\'a}mkovics} {et~al.}(2011){{\'A}d{\'a}mkovics}, {Glassgold},
  \& {Meijerink}}]{AGM11}
{{\'A}d{\'a}mkovics}, M., {Glassgold}, A.~E., \& {Meijerink}, R. 2011,
  \href{http://dx.doi.org/10.1088/0004-637X/736/2/143}{\apj, 736, 143} (AGM11)

\bibitem[{{Ag{\'u}ndez} {et~al.}(2008){Ag{\'u}ndez}, {Cernicharo}, \&
  {Goicoechea}}]{Agundez2008}
{Ag{\'u}ndez}, M., {Cernicharo}, J., \& {Goicoechea}, J.~R. 2008,
  \href{http://dx.doi.org/10.1051/0004-6361:20077927}{\aap, 483, 831}

\bibitem[{{Akimkin} {et~al.}(2013){Akimkin}, {Zhukovska}, {Wiebe}, {Semenov},
  {Pavlyuchenkov}, {Vasyunin}, {Birnstiel}, \& {Henning}}]{Akimkin2013}
{Akimkin}, V., {Zhukovska}, S., {Wiebe}, D., {et~al.} 2013,
  \href{http://dx.doi.org/10.1088/0004-637X/766/1/8}{\apj, 766, 8}

\bibitem[{{Aresu} {et~al.}(2011){Aresu}, {Kamp}, {Meijerink}, {Woitke}, {Thi},
  \& {Spaans}}]{Aresu2011}
{Aresu}, G., {Kamp}, I., {Meijerink}, R., {et~al.} 2011,
  \href{http://dx.doi.org/10.1051/0004-6361/201015449}{\aap, 526, A163}

\bibitem[{{Aresu} {et~al.}(2012){Aresu}, {Meijerink}, {Kamp}, {Spaans}, {Thi},
  \& {Woitke}}]{Aresu2012}
{Aresu}, G., {Meijerink}, R., {Kamp}, I., {et~al.} 2012,
  \href{http://dx.doi.org/10.1051/0004-6361/201219864}{\aap, 547, A69}

\bibitem[{{Baulch}(2005)}]{Baulch2005}
{Baulch}, D.~L. 2005, \href{http://dx.doi.org/10.1063/1.1748524}{Journal of
  Physical and Chemical Reference Data, 34, 757}

\bibitem[{{Bethell} \& {Bergin}(2009)}]{BB09}
{Bethell}, T., \& {Bergin}, E. 2009,
  \href{http://dx.doi.org/10.1126/science.1176879}{Science, 326, 1675} (BB09)

\bibitem[{{Bethell} \& {Bergin}(2011)}]{BB11}
{Bethell}, T.~J., \& {Bergin}, E.~A. 2011,
  \href{http://dx.doi.org/10.1088/0004-637X/739/2/78}{\apj, 739, 78}

\bibitem[{{Bruderer} {et~al.}(2012){Bruderer}, {van Dishoeck}, {Doty}, \&
  {Herczeg}}]{Bruderer2012}
{Bruderer}, S., {van Dishoeck}, E.~F., {Doty}, S.~D., \& {Herczeg}, G.~J. 2012,
  \href{http://dx.doi.org/10.1051/0004-6361/201118218}{\aap, 541, A91}

\bibitem[{{Burke} \& {Hollenbach}(1983)}]{BH83}
{Burke}, J.~R., \& {Hollenbach}, D.~J. 1983,
  \href{http://dx.doi.org/10.1086/160667}{\apj, 265, 223}

\bibitem[{{Carr} \& {Najita}(2008)}]{CN08}
{Carr}, J.~S., \& {Najita}, J.~R. 2008,
  \href{http://dx.doi.org/10.1126/science.1153807}{Science, 319, 1504}

\bibitem[{{Carr} \& {Najita}(2011)}]{CN11}
---. 2011, \href{http://dx.doi.org/10.1088/0004-637X/733/2/102}{\apj, 733, 102}

\bibitem[{{Carr} {et~al.}(2004){Carr}, {Tokunaga}, \& {Najita}}]{Carr2004}
{Carr}, J.~S., {Tokunaga}, A.~T., \& {Najita}, J. 2004,
  \href{http://dx.doi.org/10.1086/381356}{\apj, 603, 213}

\bibitem[{{Cazaux} \& {Tielens}(2004)}]{Cazaux2004}
{Cazaux}, S., \& {Tielens}, A.~G.~G.~M. 2004,
  \href{http://dx.doi.org/10.1086/381775}{\apj, 604, 222}

\bibitem[{{Cazaux} \& {Tielens}(2010)}]{CT2010}
{Cazaux}, S., \& {Tielens}, A.~G.~G.~M. 2010,
  \href{http://dx.doi.org/10.1088/0004-637X/715/1/698}{\apj, 715, 698}

\bibitem[{{Ciesla} \& {Cuzzi}(2006)}]{Ciesla2006}
{Ciesla}, F.~J., \& {Cuzzi}, J.~N. 2006,
  \href{http://dx.doi.org/10.1016/j.icarus.2005.11.009}{\icarus, 181, 178}

\bibitem[{{Crovisier}(1989)}]{Crovisier1989}
{Crovisier}, J. 1989, \aap, 213, 459

\bibitem[{{D'Alessio} {et~al.}(1999){D'Alessio}, {Calvet}, {Hartmann},
  {Lizano}, \& {Cant{\'o}}}]{DAlessio1999}
{D'Alessio}, P., {Calvet}, N., {Hartmann}, L., {Lizano}, S., \& {Cant{\'o}}, J.
  1999, \href{http://dx.doi.org/10.1086/308103}{\apj, 527, 893}

\bibitem[{{Dullemond} \& {Monnier}(2010)}]{Dullemond2010}
{Dullemond}, C.~P., \& {Monnier}, J.~D. 2010,
  \href{http://dx.doi.org/10.1146/annurev-astro-081309-130932}{\araa, 48, 205}

\bibitem[{{Fogel} {et~al.}(2011){Fogel}, {Bethell}, {Bergin}, {Calvet}, \&
  {Semenov}}]{Fogel2011}
{Fogel}, J.~K.~J., {Bethell}, T.~J., {Bergin}, E.~A., {Calvet}, N., \&
  {Semenov}, D. 2011, \href{http://dx.doi.org/10.1088/0004-637X/726/1/29}{\apj,
  726, 29}

\bibitem[{{France} {et~al.}(2011){France}, {Yang}, \& {Linsky}}]{France2011}
{France}, K., {Yang}, H., \& {Linsky}, J.~L. 2011,
  \href{http://dx.doi.org/10.1088/0004-637X/729/1/7}{\apj, 729, 7}

\bibitem[{{France} {et~al.}(2012){France}, {Schindhelm},
  {Herczeg}, {Brown}, {Abgrall}, {Alexander}, {Bergin}, {Brown}, {Linsky},
  {Roueff}, \& {Yang}}]{France2012a}
{France}, K., {Schindhelm}, E., {Herczeg}, G.~J., {et~al.} 2012,
  \href{http://dx.doi.org/10.1088/0004-637X/756/2/171}{\apj, 756, 171}

\bibitem[{{Garmire} {et~al.}(2000){Garmire}, {Feigelson}, {Broos},
  {Hillenbrand}, {Pravdo}, {Townsley}, \& {Tsuboi}}]{Garmire2000}
{Garmire}, G., {Feigelson}, E.~D., {Broos}, P., {et~al.} 2000,
  \href{http://dx.doi.org/10.1086/301523}{\aj, 120, 1426}

\bibitem[{{Glassgold} {et~al.}(2012){Glassgold}, {Galli}, \&
  {Padovani}}]{GGP12}
{Glassgold}, A.~E., {Galli}, D., \& {Padovani}, M. 2012,
  \href{http://dx.doi.org/10.1088/0004-637X/756/2/157}{\apj, 756, 157}

\bibitem[{{Glassgold} {et~al.}(2009){Glassgold}, {Meijerink}, \&
  {Najita}}]{GMN09}
{Glassgold}, A.~E., {Meijerink}, R., \& {Najita}, J.~R. 2009,
  \href{http://dx.doi.org/10.1088/0004-637X/701/1/142}{\apj, 701, 142} (GMN09)

\bibitem[{{Glassgold} {et~al.}(2004){Glassgold}, {Najita}, \& {Igea}}]{GNI04}
{Glassgold}, A.~E., {Najita}, J., \& {Igea}, J. 2004,
  \href{http://dx.doi.org/10.1086/424509}{\apj, 615, 972} (GNI04)

\bibitem[{{Gorti} \& {Hollenbach}(2008)}]{Gorti2008}
{Gorti}, U., \& {Hollenbach}, D. 2008,
  \href{http://dx.doi.org/10.1086/589616}{\apj, 683, 287}

\bibitem[{{G{\"u}del} {et~al.}(2010){G{\"u}del}, {Lahuis}, {Briggs}, {Carr},
  {Glassgold}, {Henning}, {Najita}, {van Boekel}, \& {van
  Dishoeck}}]{Gudel2010}
{G{\"u}del}, M., {Lahuis}, F., {Briggs}, K.~R., {et~al.} 2010,
  \href{http://dx.doi.org/10.1051/0004-6361/200913971}{\aap, 519, A113}

\bibitem[{{Habing}(1968)}]{Habing1968}
{Habing}, H.~J. 1968, \bain, 19, 421

\bibitem[{{Heinzeller} {et~al.}(2011){Heinzeller}, {Nomura}, {Walsh}, \&
  {Millar}}]{Heinzeller2011}
{Heinzeller}, D., {Nomura}, H., {Walsh}, C., \& {Millar}, T.~J. 2011,
  \href{http://dx.doi.org/10.1088/0004-637X/731/2/115}{\apj, 731, 115}

\bibitem[{{Herczeg} {et~al.}(2004){Herczeg}, {Wood}, {Linsky}, {Valenti}, \&
  {Johns-Krull}}]{Herczeg2004}
{Herczeg}, G.~J., {Wood}, B.~E., {Linsky}, J.~L., {Valenti}, J.~A., \&
  {Johns-Krull}, C.~M. 2004, \href{http://dx.doi.org/10.1086/383340}{\apj, 607,
  369}
\bibitem[{{Hirose} \& {Turner}(2011)}]{HiroseTurner2011}
{Hirose}, S., \& {Turner}, N.~J. 2011,
  \href{http://dx.doi.org/10.1088/2041-8205/732/2/L30}{\apjl, 732, L30}

\bibitem[{{Kamp} \& {Dullemond}(2004)}]{Kamp2004}
{Kamp}, I., \& {Dullemond}, C.~P. 2004,
  \href{http://dx.doi.org/10.1086/424703}{\apj, 615, 991}

\bibitem[{{Kamp} {et~al.}(2010){Kamp}, {Tilling}, {Woitke}, {Thi}, \&
  {Hogerheijde}}]{Kamp2010}
{Kamp}, I., {Tilling}, I., {Woitke}, P., {Thi}, W.-F., \& {Hogerheijde}, M.
  2010, \href{http://dx.doi.org/10.1051/0004-6361/200913076}{\aap, 510, A18}

\bibitem[{{Kamp} {et~al.}(2011){Kamp}, {Woitke}, {Pinte}, {Tilling}, {Thi},
  {Menard}, {Duchene}, \& {Augereau}}]{Kamp2011}
{Kamp}, I., {Woitke}, P., {Pinte}, C., {et~al.} 2011,
  \href{http://dx.doi.org/10.1051/0004-6361/201016399}{\aap, 532, A85}

\bibitem[{{Lemaire} {et~al.}(2010){Lemaire}, {Vidali}, {Baouche}, {Chehrouri},
  {Chaabouni}, \& {Mokrane}}]{Lemaire2010}
{Lemaire}, J.~L., {Vidali}, G., {Baouche}, S., {et~al.} 2010,
  \href{http://dx.doi.org/10.1088/2041-8205/725/2/L156}{\apjl, 725, L156}

\bibitem[{{Markwick} {et~al.}(2002){Markwick}, {Ilgner}, {Millar}, \&
  {Henning}}]{Markwick2002}
{Markwick}, A.~J., {Ilgner}, M., {Millar}, T.~J., \& {Henning}, T. 2002,
  \href{http://dx.doi.org/10.1051/0004-6361:20020050}{\aap, 385, 632}

\bibitem[{{Mathis} {et~al.}(1977){Mathis}, {Rumpl}, \& {Nordsieck}}]{MRN}
{Mathis}, J.~S., {Rumpl}, W., \& {Nordsieck}, K.~H. 1977,
  \href{http://dx.doi.org/10.1086/155591}{\apj, 217, 425}

\bibitem[{{Meijerink} {et~al.}(2012){Meijerink}, {Aresu}, {Kamp}, {Spaans},
  {Thi}, \& {Woitke}}]{Meijerink2012}
{Meijerink}, R., {Aresu}, G., {Kamp}, I., {et~al.} 2012,
  \href{http://dx.doi.org/10.1051/0004-6361/201219863}{\aap, 547, A68}

\bibitem[{{Mota} {et~al.}(2005){Mota}, {Parafita}, {Giuliani},
  {Hubin-Franskin}, {Lourenco}, {Garcia}, {Hoffmann}, {Mason}, {Ribeiro},
  {Raposo}, \& {Limao-Vieira}}]{Mota2005}
{Mota}, R., {Parafita}, R., {Giuliani}, A., {et~al.} 2005,
  \href{http://dx.doi.org/10.1016/j.cplett.2005.09.073}{Chemical Physics
  Letters, 416, 152}

\bibitem[{{Najita} {et~al.}(2011){Najita}, {{\'A}d{\'a}mkovics}, \&
  {Glassgold}}]{NAG11}
{Najita}, J.~R., {{\'A}d{\'a}mkovics}, M., \& {Glassgold}, A.~E. 2011,
  \href{http://dx.doi.org/10.1088/0004-637X/743/2/147}{\apj, 743, 147} (NAG11)

\bibitem[{{Najita} {et~al.}(2013){Najita}, {Carr}, {Pontoppidan}, {Salyk}, {van
  Dishoeck}, \& {Blake}}]{Najita2013}
{Najita}, J.~R., {Carr}, J.~S., {Pontoppidan}, K.~M., {et~al.} 2013,
  \href{http://dx.doi.org/10.1088/0004-637X/766/2/134}{\apj, 766, 134}

\bibitem[{{Najita} {et~al.}(2010){Najita}, {Carr}, {Strom}, {Watson},
  {Pascucci}, {Hollenbach}, {Gorti}, \& {Keller}}]{Najita2010}
{Najita}, J.~R., {Carr}, J.~S., {Strom}, S.~E., {Watson}, D.~M., {Pascucci},
  I., {Hollenbach}, D., {Gorti}, U., \& {Keller}, L. 2010, \apj, 712, 274

\bibitem[{{Nee} \& {Lee}(1984)}]{NeeLee1984}
{Nee}, J.~B., \& {Lee}, L.~C. 1984,
  \href{http://dx.doi.org/10.1063/1.447387}{\jcp, 81, 31}

\bibitem[{{Neufeld} \& {Kaufman}(1993)}]{NK93}
{Neufeld}, D.~A., \& {Kaufman}, M.~J. 1993,
  \href{http://dx.doi.org/10.1086/173388}{\apj, 418, 263}

\bibitem[{{Nomura} {et~al.}(2009){Nomura}, {Aikawa}, {Nakagawa}, \&
  {Millar}}]{Nomura2009}
{Nomura}, H., {Aikawa}, Y., {Nakagawa}, Y., \& {Millar}, T.~J. 2009,
  \href{http://dx.doi.org/10.1051/0004-6361:200810206}{\aap, 495, 183}

\bibitem[{{Nomura} {et~al.}(2007){Nomura}, {Aikawa}, {Tsujimoto}, {Nakagawa},
  \& {Millar}}]{Nomura2007}
{Nomura}, H., {Aikawa}, Y., {Tsujimoto}, M., {Nakagawa}, Y., \& {Millar}, T.~J.
  2007, \href{http://dx.doi.org/10.1086/513419}{\apj, 661, 334}

\bibitem[{{Nomura} \& {Millar}(2005)}]{Nomura2005}
{Nomura}, H., \& {Millar}, T.~J. 2005,
  \href{http://dx.doi.org/10.1051/0004-6361:20052809}{\aap, 438, 923}

\bibitem[{{Pascucci} {et~al.}(2009){Pascucci}, {Apai}, {Luhman}, {Henning},
  {Bouwman}, {Meyer}, {Lahuis}, \& {Natta}}]{Pascucci2009}
{Pascucci}, I., {Apai}, D., {Luhman}, K., {et~al.} 2009,
  \href{http://dx.doi.org/10.1088/0004-637X/696/1/143}{\apj, 696, 143}

\bibitem[{{Pontoppidan} {et~al.}(2010{\natexlab{a}}){Pontoppidan}, {Salyk},
  {Blake}, \& {K{\"a}ufl}}]{Pontoppidan2010a}
{Pontoppidan}, K.~M., {Salyk}, C., {Blake}, G.~A., \& {K{\"a}ufl}, H.~U.
  2010{\natexlab{a}}, \apjl, 722, L173

\bibitem[{{Pontoppidan} {et~al.}(2010{\natexlab{b}}){Pontoppidan}, {Salyk},
  {Blake}, {Meijerink}, {Carr}, \& {Najita}}]{Pontoppidan2010b}
{Pontoppidan}, K.~M., {Salyk}, C., {Blake}, G.~A., {Meijerink}, R., {Carr},
  J.~S., \& {Najita}, J. 2010{\natexlab{b}}, \apj, 720, 887

\bibitem[{{Rodgers} \& {Charnley}(2005)}]{RodgersCharnley2005}
{Rodgers}, S.~D. \& {Charnley}, S.~B. 2005, \mnras, 356, 1542

\bibitem[{{Roser} {et~al.}(2003){Roser}, {Swords}, {Vidali}, {Manic{\`o}}, \&
  {Pirronello}}]{Roser2003}
{Roser}, J.~E., {Swords}, S., {Vidali}, G., {Manic{\`o}}, G., \& {Pirronello},
  V. 2003, \href{http://dx.doi.org/10.1086/379095}{\apjl, 596, L55}

\bibitem[{{Salyk} {et~al.}(2011){Salyk}, {Pontoppidan}, {Blake}, {Najita}, \&
  {Carr}}]{Salyk2011}
{Salyk}, C., {Pontoppidan}, K.~M., {Blake}, G.~A., {Najita}, J.~R., \& {Carr},
  J.~S. 2011, \href{http://dx.doi.org/10.1088/0004-637X/731/2/130}{\apj, 731,
  130}

\bibitem[{{Schindhelm} {et~al.}(2012){Schindhelm}, {France}, {Herczeg},
  {Bergin}, {Yang}, {Brown}, {Brown}, {Linsky}, \& {Valenti}}]{Schindhelm2012}
{Schindhelm}, E., {France}, K., {Herczeg}, G.~J., {et~al.} 2012,
  \href{http://dx.doi.org/10.1088/2041-8205/756/1/L23}{\apjl, 756, L23}

\bibitem[{{Shakura} \& {Sunyaev}(1973)}]{SS73}
{Shakura}, N.~I., \& {Sunyaev}, R.~A. 1973, \aap, 24, 337

\bibitem[{{Sizun} {et~al.}(2010){Sizun}, {Bachellerie}, {Aguillon}, \&
  {Sidis}}]{Sizun2010}
{Sizun}, M., {Bachellerie}, D., {Aguillon}, F., \& {Sidis}, V. 2010,
  \href{http://dx.doi.org/10.1016/j.cplett.2010.08.039}{Chemical Physics
  Letters, 498, 32}

\bibitem[{{Spitzer} \& {Cochran}(1973)}]{Spitzer73}
{Spitzer}, Jr., L., \& {Cochran}, W.~D. 1973,
  \href{http://dx.doi.org/10.1086/181349}{\apjl, 186, L23}

\bibitem[{{Thi} {et~al.}(2010){Thi}, {Woitke}, \& {Kamp}}]{Thi2010}
{Thi}, W.-F., {Woitke}, P., \& {Kamp}, I. 2010,
  \href{http://dx.doi.org/10.1111/j.1365-2966.2009.16162.x}{\mnras, 407, 232}

\bibitem[{{Tielens}(2010)}]{Tielens2010}
{Tielens}, A.~G.~G.~M. 2010, {The Physics and Chemistry of the rstellar
  Medium} (Cambridge University Press)

\bibitem[{{van Dishoeck} \& {Dalgarno}(1984)}]{vDD84}
{van Dishoeck}, E.~F., \& {Dalgarno}, A. 1984,
  \href{http://dx.doi.org/10.1086/161729}{\apj, 277, 576}

\bibitem[{{Walsh} {et~al.}(2012){Walsh}, {Nomura}, {Millar}, \&
  {Aikawa}}]{Walsh2012}
{Walsh}, C., {Nomura}, H., {Millar}, T.~J., \& {Aikawa}, Y. 2012,
  \href{http://dx.doi.org/10.1088/0004-637X/747/2/114}{\apj, 747, 114}

\bibitem[{{Weingartner} \& {Draine}(2001)}]{WD01}
{Weingartner}, J.~C., \& {Draine}, B.~T. 2001,
  \href{http://dx.doi.org/10.1086/320852}{\apjs, 134, 263}

\bibitem[{{Williams} \& {Cieza}(2011)}]{Williams2011}
{Williams}, J.~P., \& {Cieza}, L.~A. 2011,
  \href{http://dx.doi.org/10.1146/annurev-astro-081710-102548}{\araa, 49, 67}

\bibitem[{{Woitke} {et~al.}(2009{\natexlab{a}}){Woitke}, {Kamp}, \&
  {Thi}}]{Woitke2009a}
{Woitke}, P., {Kamp}, I., \& {Thi}, W.-F. 2009{\natexlab{a}},
  \href{http://dx.doi.org/10.1051/0004-6361/200911821}{\aap, 501, 383}

\bibitem[{{Woitke} {et~al.}(2009{\natexlab{b}}){Woitke}, {Thi}, {Kamp}, \&
  {Hogerheijde}}]{Woitke2009b}
{Woitke}, P., {Thi}, W.-F., {Kamp}, I., \& {Hogerheijde}, M.~R.
  2009{\natexlab{b}},
  \href{http://dx.doi.org/10.1051/0004-6361/200912249}{\aap, 501, L5}

\bibitem[{{Woods} \& {Willacy}(2009)}]{Woods2009}
{Woods}, P.~M., \& {Willacy}, K. 2009,
  \href{http://dx.doi.org/10.1088/0004-637X/693/2/1360}{\apj, 693, 1360}

\bibitem[{{Yang} {et~al.}(2012){Yang}, {Herczeg}, {Linsky}, {Brown},
  {Johns-Krull}, {Ingleby}, {Calvet}, {Bergin}, \& {Valenti}}]{Yang2012}
{Yang}, H., {Herczeg}, G.~J., {Linsky}, J.~L., {et~al.} 2012,
  \href{http://dx.doi.org/10.1088/0004-637X/744/2/121}{\apj, 744, 121}

\end{thebibliography}
\end{document}